\documentclass[tikz,12pt]{article}
\usepackage{mheckII}

\usepackage[T1]{fontenc}

\usepackage{tikz-feynman}

\usepackage{manfnt}
\usepackage{longtable}

\usepackage{fancybox}

\usepackage{multirow}

\usepackage{blkarray}

\usepackage{tikz} 
\usetikzlibrary{matrix}

\theoremstyle{theorem}
\newtheorem*{thm}{Theorem}

\newtheorem{fact}{Fact}

\newtheorem{corl}{Corollary}
\newtheorem{pro}{Proposition}

\newtheorem{lem}{Lemma}
\newtheorem*{lem*}{Lemma}

\newtheorem{que}{Question}

\newtheorem*{ans}{Answer}

\theoremstyle{definition}

\newtheorem{rem}{Remark}

\newtheorem{exe}{Example}[section]

\def\Dsl{\,\raise.15ex\hbox{/}\mkern-13.5mu D}
\def\dsl{\,\raise.25ex\hbox{/}\mkern-10.5mu \partial}

\title{The Weil Correspondence and\\ Universal Special Geometry 
}

\authors{Sergio Cecotti\footnote{e-mail: {\tt cecotti@bimsa.cn}}\vskip 9pt

\centerline{Beijing Institute of Mathematical Sciences and Applications (BIMSA)}
\centerline{Huaibei Town, Huairou District, Beijing 101408, China}
\centerline{and}
\centerline{Qiuzhen College, Tsinghua University, Beijing, China,}
\centerline{SISSA,
via Bonomea 265, Trieste, Italy}
}

\abstract{The Weil correspondence states that the datum of a Seiberg-Witten differential is equivalent to an algebraic group extension of the integrable system associated to the Seiberg-Witten geometry.
Remarkably this group extension represents quantum consistent couplings for the $\cn=2$ QFT if and only if the extension is \emph{anti-affine} in the algebro-geometric sense. The \emph{universal special geometry} is the algebraic integrable system whose Lagrangian fibers are the anti-affine extension groups; it is defined over a base $\mathscr{B}$ parametrized by the Coulomb coordinates and the couplings. On the total space of the universal geometry there is a canonical (holomorphic) Euler differential. The ordinary Seiberg-Witten geometries
at fixed couplings are symplectic quotients of the universal one, and the Seiberg-Witten differential
arises as the reduction of the Euler one in accordance with the Weil correspondence. This universal viewpoint allows to study geometrically the flavor symmetry of the $\cn=2$ SCFT in terms of the 
Mordell-Weil lattice (with N\'eron-Tate height) of the Albanese variety $A_\mathbb{L}$ of the universal geometry seen as a quasi-Abelian variety $Y_\mathbb{L}$ defined over the function field $\mathbb{L}\equiv\C(\mathscr{B})$.}

\begin{document}
\maketitle

\tableofcontents

\newpage 

\section{Introduction and Overview}

In this paper ``special geometry''
stands for the holomorphic integrable system\footnote{\ A holomorphic integral system is an even-dimensional complex manifold $\mathscr{X}$ equipped with a non-degenerate, holomorphic, closed $(2,0)$-form $\Omega$. In our set-up $\mathscr{X}$ is in addition \emph{algebraic.}} $(\mathscr{X},\Omega)$
associated to the Seiberg-Witten geometry of a 4d $\cn=2$ QFT \cite{SW1,SW2,donagi0,donagi1}. To avoid
special cases (which require a slightly different treatment)
we make once and for all the assumption that \textit{no subsector of the QFT has a weakly coupled
Lagrangian formulation.}\footnote{\ This is a ``rigidity'' condition. It is equivalent to the requirement that the geometry is a deformation of a $\C^\times$-isoinvariant special geometry with no $\C^\times$-isoinvariant deformations.} 
The basic message of this paper is that the traditional approach to special geometry does not completely capture the richness of the theory, and fails to address several issues,
while there is a more elegant geometric framework where these questions are fully dealt with (at least in principle).

\medskip

There is an on-going program \cite{A1,A2,A3,A4,A5,A6,A7,A8,A9,A10,A11,A12,A13,A14,A15,A16,A17,A18,A19,A20,A21,A22,A23} to classify all existing $\cn=2$ SCFT
using special geometry. The holy grail of the program is to construct \emph{all} $\cn=2$ SCFTs
with no stringy or Lagrangian construction (or rule out their existence).
While partially successful, this geometric program suffers from a serious drawback: to a given
scale-invariant special geometry there correspond several inequivalent SCFTs with distinct flavor
symmetry groups, central charges, etc. To complete the program one needs
 a \emph{geometric} understanding of the set of SCFTs described by a 
 a given scale-invariant special geometry. The traditional viewpoint gives no clue on this issue. There are many other features of SCFTs with no geometric intepretation in that set-up. To make a simple example:
 from QFT we know that the relevant chiral
deformations of a SCFT are in one-to-one correspondence with the elements of the chiral ring
of dimension $1<\Delta<2$. How do we understand this elementary fact in \emph{purely geometric} terms?
We also know that the mass deformations span the Cartan subalgebra of the flavor symmetry group $F$.
What is the \emph{geometric} interpretation of these deformations, and how we may use them to determine \emph{geometrically} the allowed $F$'s? In this direction an open problem is to determine the maximal  rank $f_\text{max}$ of the flavor group for a rank-$r$ SCFT. What is $f_\text{max}$ geometrically?

\medskip

Going to details, there are many aspects of the standard story
which don't look fully satisfactory. One is the Seiberg-Witten differential $\lambda$. The traditional viewpoint is that $\lambda$ is a meromorphic differential
on $\mathscr{X}$
whose periods are the (local) special coordinates. This looks not to be the complete story on two counts:
\begin{itemize}
\item[(a)] for a SCFT $\lambda$ is more properly an \emph{infinitesimal
symmetry} of the special geometry $\mathscr{X}$. An infinitesimal symmetry of $\mathscr{X}$
is given by a holomorphic vector field $\ce\in \mathfrak{aut}(\mathscr{X})$. In a symplectic manifold 
vectors and 1-forms are equivalent, and we can write our infinitesimal symmetry as the 1-form $\lambda\equiv \iota_\ce \Omega$,
hiding its role as the generator of the connected component of $\mathsf{Aut}(\mathscr{X})$.\footnote{\ We stress that since $\lambda$ is the generator of $\mathsf{Aut}(\mathscr{X})^0$, it 
is canonically determined by the underlying symplectic manifold $(\mathscr{X},\Omega)$ and not an additional datum.}
 Away from conformality the relation of $\lambda$ with the geometric symmetries seems to be lost, while one suspects that it must hold in general \emph{in the appropriate sense}.
\item[(b)] In general one has\footnote{\ This equation is meant in the sense of cohomology of currents. Note that the $(2,0)$ form $\Omega$ represents a $(1,1)$ class not a $(2,0)$ one.} 
\begin{equation}
[\Omega]=[d\lambda]=\sum_a m_a D_a
\end{equation}
where the $D_a$ are polar divisors of $\lambda$ and the coefficients (residues) $m_a$ are mass parameters.
 It has been observed by Ron Donagi \cite{donagi1} that this expression agrees with the
Duistermaat-Heckman (DH) formula \cite{DH,cannas}: the cohomology class of the symplectic form depends linearly on the values of the momentum map. Now, while the parameters $m_a$ are, in a broad sense, momentum maps, the DH
theorem refers to a particular geometric situation namely the Marsden-Weinstein-Meyer (MWM) symplectic quotient \cite{cannas}. In the usual formulation of special geometry the manifold $\mathscr{X}$ is \emph{not} a MWM quotient of some bigger integrable system, so something looks missing from the picture.  
\end{itemize}

Taking seriously Donagi's suggestion, one concludes that there should exist a \emph{much bigger integrable system} in the form of a \emph{universal special geometry} $(\mathscr{Y},\boldsymbol{\Omega})$ of which the ordinary one $(\mathscr{X},\Omega)$ is  a symplectic reduction. The existence of the universal geometry also follows from standard physical considerations, see the paragraph below
about the physical interpretation via \emph{spurions}.
In the spirit of item (a) one expects that in the universal geometry the differential $\boldsymbol{\lambda}$
(or rather its dual vector field $\ce$) generates holomorphic automorphisms of $\mathscr{Y}$, so
$\boldsymbol{\lambda}$ must be canonically determined by the geometry of $\mathcal{Y}$. In particular the polar classes
$[D_a]$ are uniquely fixed by $(\mathscr{Y},\boldsymbol{\Omega})$. Since we can read
the flavor symmetry group $F$ from the $[D_a]$'s, going to the universal geometry $(\mathscr{Y},\boldsymbol{\Omega})$ should also
solve (in principle) the issue of a geometric
understanding of flavor symmetry.

The story is a bit more complicated since, in addition to the mass deformations (associated to the flavor symmetry), we have the \emph{relevant} couplings that also require a geometric interpretation.
They should also be determined
by the universal symmetry $\ce\in\mathfrak{aut}(\mathscr{Y})$ of the bigger geometric framework.
In other words, there must be an ``universal'' geometric theory which encompasses \emph{all kinds} of physically allowed interactions
and singles out the ones which are consistent at the full quantum level. 

\medskip

Our proposal for the solution of these issues (and many others) is based on an old mathematical gadget: the ``Weil correspondence'' that we shall review in a moment after some preliminary.

\paragraph{The fields of definition $\mathbb{K}$ and $\mathbb{L}$.}
We make the observation that an \emph{ordinary} special geometry $\mathscr{X}$ is, in particular, 
a complex model of an Abelian variety (or scheme) $X_\mathbb{K}$ defined over the field $\mathbb{K}=\C(u_1,\dots,u_r)$
of rational functions in $r$ variables ($r$ being the \emph{rank} i.e.\! the dimension of the ordinary Coulomb branch $\mathscr{C}$). This assertion contains a mild but crucial assumption that we now make explicit.
In all known $\cn=2$ SCFT the Coulomb branch\footnote{\ Here we look at the Coulomb branch as a mere complex manifold, stripped of all other structures such as metric etc.}
$\mathscr{C}$ is a copy of $\C^r$. However, as pointed out in ref.\!\cite{A8}, there are reasons to believe that more general Coulomb branches are also allowed. Since the Coulomb branch $\mathscr{C}$
is the spectrum of the chiral ring $\mathscr{R}$, 
in any quantum-consistent $\cn=2$ SCFT $\mathscr{C}$  is
an irreducible, reduced, normal, affine (complex) variety with an algebraic $\C^\times$-action (see section 2 for more details). 
\begin{que} {Which (possibly singular) affine varieties are the Coulomb branch $\mathscr{C}$ of some physically sound $\cn=2$ QFT?}
\end{que}

The allowed $\mathscr{C}$'s are severely restricted. If you look to a reasonable-looking candidate Coulomb branch $\not\simeq\C^r$ you almost certainly end up with the conclusion that the dimension of the deformation space of the special geometry is not equal to the number of chiral operators of dimension $\leq2$ as required
by QFT (and true for $\mathscr{C}\simeq \C^r$ \cite{A23}). The ``experimental'' evidence indicate that in each dimension $r$
there are few putative Coulomb branches
consistent with this constraint from the dimension of the deformation space; all of them are expected to have the same field of functions\footnote{\ In particular in rank-$1$ there is only one possible Coulomb branch: the plane $\C$ \cite{A11}.} $\C(\mathscr{C})\equiv \C(u_1,\dots,u_r)$. In this paper we make this empirical suggestion into an assumption.
Under this assumption, while the geometry of the Coulomb branch may be rather subtle, the subtleties are confined in codimension at least two,\footnote{\ The singularities, if present, are in codimension $\geq2$ since we are free to assume $\mathscr{C}$ to be normal by replacing the chiral ring $\mathscr{R}$ with its normalization, a procedure which is also suggested by QFT considerations.} while the aspects we are interested in (at this early stage of the program) are typically codimension-1 phenomena.

Since we are mainly interested in codimension-1 issues, we don't bother which specific model of the Abelian variety $X_\mathbb{K}$ yields the actual special geometry $\mathscr{X}$. At this stage we can work with $X_\mathbb{K}$
which is simply a commutative group (defined on the slightly fancier field $\mathbb{K}$) and get a lot of mileage from algebraic group theory \cite{milne}.

However the focus of this paper is not the usual special geometry $\mathscr{X}$,
 but the much bigger universal integrable system $\mathscr{Y}$. Just as we may replace $\mathscr{X}$
 with the underlying commutative group $X_\mathbb{K}$ defined over the function field $\mathbb{K}=\C(\mathscr{C})$, we may replace $\mathscr{Y}$ by a commutative algebraic group $Y_\mathbb{L}$
 defined over the bigger function field $\mathbb{L}=\C(\mathscr{B})$ which is an extension of
 $\mathbb{K}$ of transcendental degree equal to the complex dimension of the space of mass and relevant couplings.  The group-(scheme) $Y_\mathbb{L}$ is our main object of interest.

\paragraph{The Seiberg-Witten differential $\lambda$.}
In the algebraic group language the Seiberg-Witten differential $\lambda$ is a differential on $X_\mathbb{K}$
defined over $\mathbb{K}$. From the QFT point of view $\lambda$ depends on three 
distinct kinds of variables:\footnote{\  There are no marginal couplings under our
assumption that theory has no Lagrangian subsectors.}
the Coulomb coordinates $(u_1,\dots,u_r)$,
the chiral couplings $(t_1,\dots,t_k)$,
and the mass parameters $(m_1,\dots,m_f)$. The geometric distinction between the three kinds of parameters is as follows:
\begin{equation}\label{juyuqww123}
\begin{aligned}
\frac{\partial\lambda}{\partial u_i} &\in\Big\{\text{differentials of the first kind on }X_\mathbb{K}\Big\}\\
\frac{\partial\lambda}{\partial t_j} &\in\frac{\Big\{\text{differentials of the second kind on }X_\mathbb{K}\Big\}}{\Big\{\text{first kind $+$ exact differentials}\Big\}}\\
\frac{\partial\lambda}{\partial m_a} &\in\frac{\Big\{\text{differentials of the third kind on }X_\mathbb{K}\Big\}}{\Big\{\text{second kind $+$ exact differentials}\Big\}}
\end{aligned}
\end{equation}
The parameters $u_i$, $t_j$ and $m_a$ exhaust the supply of differentials, so in $\lambda$ there is no room for couplings of fancier kinds.
This is consistent with the superspace construction of $\cn=2$ QFT. 
We shall see below that, in addition, only chiral couplings consistent with quantum UV-completeness
are allowed in special geometry.

Eq.\eqref{juyuqww123} suggests that we should treat the parameters $u_i$, $t_j$ and $m_a$
on the same footing. Therefore we introduce the \emph{universal} branch $\mathscr{B}\simeq \mathscr{C}\times\mathscr{P}$
with global\footnote{\ To make the masses $m_a$ into global coordinates we may need to go to
a finite cover of the physical coupling space.} coordinates $u_i$, $t_j$, $m_a$.
We write $\mathbb{L}=\C(\mathscr{B})\simeq\C(u_i,t_j,m_a)$ for the rational field in $r+k+f$
variables. Over $\mathscr{B}$ we have the universal Abelian variety
\be
\xymatrix{\mathscr{A}\ar[r]^(0.3)\varpi &\mathscr{B}\simeq \mathscr{C}\times \mathscr{P}},
\ee
i.e.\! the family of ordinary special geometries parametrized by the coupling space $\mathscr{P}$.
Again, $\mathscr{A}\to \mathscr{B}$ is a model of an Abelian variety $A_\mathbb{L}$ defined over $\mathbb{L}$.
The Seiberg-Witten differential $\lambda$ is then a third-kind differential on $A_\mathbb{L}$
defined over $\mathbb{L}$.
 The ordinary special geometry at fixed couplings $p\equiv(\underline{t}_j, \underline{m}_a)\in\mathscr{P}$ is $\mathscr{X}\equiv\mathscr{A}|_{\mathscr{C}\times p}$.

\medskip

\paragraph{The Weil correspondence.} There are many math gadgets known under the name ``Weil correspondence''. The one of interest in this paper is perhaps less known. To put it in the proper perspective, we quote from the book by Mazur and Messing \cite{MM}:\footnote{\ The references in the quote 
insides the oval box are: Weil paper \cite{Weil}; Barsotti paper \cite{barsotti}, Serre paper \cite{serre1}
and Serre book \cite{serre2}.} 
\vskip12pt

\noindent\Ovalbox{\begin{minipage}{460pt}
In [27] Weil observed that when working on abelian variety $A$ over an arbitrary field, considerations of \emph{extensions of $A$ by a vector group} replaces the study of \emph{differentials of the second kind}, while considerations of \emph{extensions of $A$ by a torus} replaces the study of \emph{differentials of the third kind}.\\ He attributes these ideas (in the classical case) to Severi.\\ Over $\mathbb{C}$, Barsotti in [1 bis] established algebraically the isomorphism\vskip-6pt
$$
\mathsf{Ext}(A,\mathbb{G}_a)\cong \frac{\text{differentials of second kind}}{\text{holomorphic differentials}+\text{exact differentials}}
$$
(See Serre's [24] and [25] for a beautiful account of these ideas)
\end{minipage}}

\vskip16pt

Comparing this quotation with eq.\eqref{juyuqww123} we conclude that we may identify 
the couplings allowed by $\cn=2$ supersymmetry 
with commutative algebraic groups. In particular the \emph{ordinary} Seiberg-Witten geometry $(\mathscr{X},\lambda)$ (with mass and relevant deformations switched on)
defines (and is fully determined by) an algebraic group $Z_\mathbb{K}$
 over $\mathbb{K}$.
To physicists \emph{group} means
\emph{symmetry,} and we think the differential $\lambda$ as describing a symmetry of the geometry. 

Again, to get a more intrinsic and satisfactory picture one has to work with the appropriate commutative extension $Y_\mathbb{L}$
of the universal Abelian group-variety $A_\mathbb{L}$ defined over the field $\mathbb{L}$.
The physical meaning of $Y_\mathbb{L}$ will be clarified below.

\paragraph{Anti-affine (quasi-Abelian) groups.}
The Weil correspondence replaces the issue of a geometric description of the allowed physical couplings (in particular mass deformations) with the more geometric question of understanding the allowed symmetries.

However not all commutative algebraic groups $Y_\mathbb{L}$ are allowed for the following reasons. 
Consider the mass deformations of a SCFT. Physically they are associated to the flavor Lie group $F$: the mass parameters take value in the Cartan algebra $\mathfrak{car}(F)$ of $F$. 
The deformation space $\mathfrak{car}(F)$
comes with a number of specific structures: an action of the Weyl group, an invariant bilinear form, and
 an integral lattice. In addition, we expect that the rank of $F$ is bounded in each fixed rank $r\equiv\dim \mathscr{C}$. Generic torus extensions do not come with such structures, and their rank has no upper bound.
Hence a general extension is not allowed.

The case of relevant couplings is even sharper.
According to the Weil correspondence,
a SCFT with the relevant couplings switched on (but no mass deformation) corresponds to a group
$Y_\mathbb{L}$ which is a
vector extension of the universal Abelian variety $A_\mathbb{L}$. An Abelian variety has extensions by vector groups of arbitrary dimension. At the \emph{classical level} the $\cn=2$ SCFT has indeed infinitely many linearly independent chiral deformations, but almost all of them are \emph{inconsistent} at the quantum level because they spoil UV completeness. The deformations which are consistent at the full non-perturbative level
are in one-to-one correspondence with the chiral operators of the undeformed theory with scaling dimension $1<\Delta<2$, in particular their number is at most $r$. This leads to the

\begin{que}
{Which algebraic groups $Y_\mathbb{L}$ over $\mathbb{L}$ are the Weil correspondents
of $\cn=2$ QFTs which are fully consistent at the quantum level (UV complete)?}
\end{que}

The fact that all 4d $\cn=2$ QFT can be twisted \emph{\'a la} Witten \cite{ttf1,witten,marino} into a Topological Field Theory (TFT), 
leads to the following (see main text): 

\begin{ans} The algebraic group
$Y_\mathbb{L}$ must be \emph{anti-affine} (a.k.a.\! quasi-Abelian), that is,
\begin{equation}\label{antiaff}
\Gamma(Y_\mathbb{L},\co_{Y_{\mathbb{L}}})\simeq\mathbb{L}.
\end{equation}
\end{ans}
Here $\co_{Y_{\mathbb{L}}}$ is the structure sheaf of the algebraic variety $Y_\mathbb{L}$ defined over $\mathbb{L}$.
In this paper we check that \eqref{antiaff} \emph{exactly matches} the conditions from quantum consistency of the QFT. 

\paragraph{Universal special geometry.} Up to now we just considered (families of) ordinary special geometries, but this is clearly not the full story as the Donagi remark in item (b) indicates. 
There must be a bigger algebraic integrable system with underlying algebraic group $Y_\mathbb{L}$.
The dimension (over $\mathbb{L}$)
of $Y_\mathbb{L}$ is the total number of parameters $r+k+f$. A model over $\C$ of $Y_\mathbb{L}$ will be a
fibration 
 \begin{equation}\label{bigLag}
 \pi\colon\mathscr{Y}\to \mathscr{B}\simeq\mathscr{C}\times\mathscr{P},
 \end{equation}   
 where the fibers and the base $\mathscr{B}$  both have complex dimension $r+k+f$. We claim that 
 the total space $\mathscr{Y}$ is a ``symplectic variety'' with symplectic form $\boldsymbol{\Omega}$,
 while the fibers of $\pi$ are Lagrangian submanifolds. 
 We write ``symplectic variety'' between quotes because we do not have control on singularities in
 codimension $\geq2$ and it is possible (even expected) that for some $\cn=2$ QFT these singularities have no crepant resolution.\footnote{\ A crepant resolution is automatically a symplectic variety.} 
 
Let us make the claim more precise. The Lagrangian fibration  \eqref{bigLag} has a $\C^\times$ group
of automorphisms. Let $\ce$ be the \emph{Euler vector field,} that is, the generator of the Lie algebra
$\mathfrak{aut}(\mathscr{Y})$ normalized so that
\begin{equation}\label{tttqza}
\mathscr{L}_\ce\boldsymbol{\Omega}=\boldsymbol{\Omega}.
\end{equation} 
The dual 1-form $\boldsymbol{\lambda}=\iota_\ce\boldsymbol{\Omega}$ is then the universal differential which is holomorphic and canonically defined by the symmetries of the geometry. From \eqref{tttqza} $d\boldsymbol{\lambda}=\boldsymbol{\Omega}$.

The Lagrangian fibration \eqref{bigLag} is our \emph{universal special geometry}. The only difference with respect to an ordinary special geometry is that now the generic (smooth) fiber is a general \emph{quasi-Abelian variety} instead of an Abelian variety. 

\paragraph{Symplectic quotients and SW differentials.} The Lagrangian fibration \eqref{bigLag} is an (algebraic) Liouville integrable system
whose smooth fibers may be non-compact. Its Hamiltonians in involution are the regular functions on the affine base
$\mathscr{B}$. We can perform the symplectic quotient at fixed values $\underline{t}_j,\underline{m}_a$ of the couplings $t_j$ and masses $m_a$ setting
\begin{gather}
\mathscr{C}=\big\{(u_i,t_j,m_a)\in\mathscr{B}\colon t_j=\underline{t}_j, m_a=\underline{m}_a\big\}\subset\mathscr{B},\\ \mathscr{X}\equiv\pi^{-1}(\mathscr{C})/H\to\mathscr{C},
\end{gather}
where $H$ is the group generated by the vector fields dual to the differentials $dt_j$'s and $dm_a$'s.
The reduced fibration $\mathscr{X}\to\mathscr{C}$ is an ordinary special geometry over the ordinary Coulomb branch $\mathscr{C}$ whose
symplectic form class $[\Omega]$ depends linearly on the parameters $t_j$, $m_a$ (in facts only on the masses $m_a$) according to the Duistermaat-Heckman formula \cite{DH}.
More in detail: on $\mathscr{Y}$ we have the God given universal Euler differential $\boldsymbol{\lambda}$
which induces a differential $\lambda$ on the symplectic reduction $\mathscr{X}$
such that $d\lambda=\Omega$. The differential $\lambda$ arising from the symplectic quotient
 coincides with the usual
Seiberg-Witten differential: in this framework \textit{the Seiberg-Witten differential is constructed out of the symmetries of the problem.} The construction of $\lambda$ is very explicit:
the basic tool is Picard's construction of anti-affine groups. It is easy to check in the examples that $\lambda$ is the physically correct differential.

The reader may be puzzled. The universal differential $\boldsymbol{\lambda}$ is perfectly holomorphic. How it happens that $\lambda$ has now become \emph{meromorphic?}. The point is that all group extension is
a principal bundle, and to write explicit expressions we need to choose a gauge on this bundle. Just as for the magnetic monopole in $\R^3$, if you insist to use a single coordinate chart you are forced to pick up a \emph{singular gauge,} and the gauge connection $A$ will look singular in that gauge; but the singularity is a mere gauge artifact: in a regular gauge it looks perfectly regular.
The Seiberg-Witten differential is singular (in presence of non-trivial couplings)
just because we write it in a (convenient) singular gauge. 

\paragraph{Physical interpetation: Spurions.} Finally we give the physical motivations for the
extension of the special geometry from the ordinary version $\mathscr{X}$ to the universal one
 $\mathscr{Y}$. From the viewpoint of the superspace approach to SUSY
QFTs, we can always see the couplings as $\cn=2$ chiral superfields which are frozen to their
constant vev's. Such non-dynamical superfields are usually called \emph{spurions}.
In particular the masses may be thought of as complex scalars in SYM superfields which weakly gauge the flavor symmetry $F$: physically we may see the flavor symmetry as the zero-coupling limit of a gauge symmetry.
In other words, to switch on the mass deformations we gauge the Cartan subgroup\footnote{\ Properly speaking one has to gauge the full group $F$; however in the $\lambda\to0$ limit the difference becomes inessential.} of $F$ with $\mathrm{rank}\,F$ spurion vector supermultiplets,
adding a kinetic term for them
\begin{equation}
\frac{1}{g_f^2}\int\!\Big(-\frac{1}{4}F_{\mu\nu\, a}\,F^{\mu\nu\, a}+\partial_\mu M_a^*\,\partial^\mu M_a+\text{fermions}\Big)d^4x,
\end{equation}
and then send $g_f\to0$ while keeping fixed $\langle M_a\rangle=\underline{m}_a$.
In the same way the relevant couplings $t_j$ can be seen as $\cn=2$ chiral superfields $T_j$
which are frozen to constant values 
by rescaling their kinetic terms\footnote{\ The $T_i$'s kinetic terms are irrelevant operators; the situation is
rather similar to the case of (2,2) Landau-Ginzburg in two dimensions \cite{ttstar}.} by an overall factor  $1/\epsilon^2$ and then sending $\epsilon\to 0$
while keeping fixed $\langle T_j\rangle =\underline{t}_j$. The interaction with couplings $t_j$ is written
as an integral over the $\cn=2$ chiral superspace
\be
\sum_j\int d^4 x\,d^4\theta\, T_j\, \phi_j+\text{h.c.} \xrightarrow{\ \epsilon\to0\ } \sum_j t_j\int d^4 x\,d^4\theta\, \phi_j+\text{h.c.}
\ee
where the $\phi_j$'s are the chiral superfields whose first components are the operators $u_j\in\mathscr{R}$
of dimension $1<\Delta_j<2$.

If we add only vector spurions, at \emph{finite} $g_f$ we have just an ordinary\footnote{\ More precisely the \emph{germ} of an ordinary special geometry since after the gauging of the flavor group the theory is typically non UV-complete. Since this is a physicists' argument, we dispense the reader with too many technical pedantries.} special geometry
with a Coulomb branch $\mathscr{B}$ of dimension $r+f$. Sending $g_f\to0$ has two effects:
\begin{itemize}
\item[(1)] it friezes the Coulomb coordinates associated to the flavor group to the values $\underline{m}_a\equiv \langle M_a\rangle$,
thus replacing $\mathscr{B}$ with the  subvariety $\mathscr{C}=\{m_a=\underline{m}_a\}$ i.e.\! with the ordinary
Coulomb branch;
\item[(2)] the (generic) fiber over $\mathscr{C}$ degenerates from an Abelian variety of dimension $r+f$
to a \emph{semi}-Abelian variety ($\equiv$ a torus extension of an Abelian variety). A semi-Abelian variety arises as a semi-stable degeneration of an Abelian variety, so (in a sense) this is the mildest possible degeneration for an ordinary special geometry. 
\end{itemize}
The situation with relevant coupling is similar except that the dimension of the corresponding operator is $\Delta >1$ so the fiber degeneration is more severe and we get a vector extension (an \emph{unstable} degeneration). 

Before sending $g_f,\epsilon\to 0$ both the masses $m_a$ and the chiral couplings $t_j$
are generators of the universal chiral ring $\mathscr{Q}$. In the limit these generators
are replaced by complex numbers, and $\mathscr{Q}$ specializes to the usual chiral ring
$\mathscr{R}$. The base $\mathscr{B}$ of the extended geometry is the affine variety 
$\mathsf{Spec}\,\mathscr{Q}$.

\paragraph{Open problems.} The Weil correspondence solves many issues
but it also opens new questions. The most important one is to find the maximal rank $f(r)$
of the flavor group $F$ as a function of the dimension $r$ of the ordinary Coulomb branch.
The overall picture is roughly as follows. The projective closure of (a suitable model of)
$X_\mathbb{K}$ is expected to be a Fano variety $\cf$ (possibly singular) of dimension $2r$
with the property that $\cf\setminus K$ is symplectic  ($K$ a canonical divisor). E.g.\! for $r=1$
this condition says that $\cf$ is a rational elliptic surface \cite{A11}.
The rank of the flavor group is related to the Picard number of $\cf$. Hence
the question boils down to an optimal upper bound on the Picard numbers of Fano varieties of dimension $2r$
with $\cf\setminus K$ symplectic and only physically admissible singularities.

\paragraph{Organization of the paper.} The rest of the paper contains detailed explanations, computations, and technicalities
making the above picture more precise. In section 2 we review special geometry from our
abstract viewpoint. In section 3 we discuss the anti-affine algebraic groups, study Lagrangian-fibrations whose general fibers are anti-affine groups, and describe their physical interpretation.
In section 4 we introduce the Mordell-Weil lattices and their N\'eron-Tate height pairing. We then compare the resulting structures with the physics of flavor symmetry. In section 5 we compute the Seiberg-Witten differential using the Weil correspondence and the Picard explicit construction of quasi-Abelian groups.   

\section{General special geometry}

General special geometry may be summarized as follows.
We have a polarized, normal algebraic variety $\mathscr{Y}$ over $\C$ with a
holomorphic symplectic 2-form $\boldsymbol{\Omega}$. ``Polarized'' means equipped with an integral  class $\omega\in H^2(\mathscr{Y},\Z)\cap H^{1,1}(\mathscr{Y})$ containing a K\"ahler metric. The crucial property
is that the affinization morphism 
\begin{equation}\label{affini}
\pi\colon \mathscr{Y}\to\mathsf{Spec}\,\Gamma(\mathscr{Y},\mathscr{O}_\mathscr{Y})\equiv\mathscr{B}
\end{equation}
is a fibration with (connected) Lagrangian fibers, hence $\mathscr{Y}$ is an algebraic Liouville integrable system whose
first integrals of motion in involution are precisely \emph{all} the regular functions on the `phase space' $\mathscr{Y}$. The (holomorphic) Hamiltonian vector fields
generate an action $G\curvearrowright\mathscr{Y}$
of a connected commutative group $G$ and $\pi$ is 
its momentum map. Thus
\begin{quote}\it The relevant geometries are (in particular) algebraic integrable systems whose
momentum map $\mu$ coincides with the affinization morphism $\pi$.
\end{quote}
We write $\mathscr{Q}\equiv\Gamma(\mathscr{Y},\mathscr{O}_\mathscr{Y})$ and call it the \emph{universal ring}. The fiber $\mathscr{Y}_b$ over a \emph{generic} point $b\in\mathscr{B}$
is smooth. The locus of points $b\in\mathscr{B}$ with non-smooth fiber, iff non-empty, is a divisor $\mathscr{D}\subset\mathscr{B}$ called the \emph{discriminant}. 
When $\pi$ has a section, $e\colon \mathscr{B}\to \mathscr{Y}$,
the smooth locus of a fiber $\mathring{\mathscr{Y}}_b\subset \mathscr{Y}_b$ is an algebraic group
whose connected component has the form $G/G_{e(b)}$, where $G_{e(b)}\subset G$ is the discrete isotropy subgroup. $e(b)$ is then the neutral element of the group $\mathring{\mathscr{Y}}_b^0$, and $e$ is called the \emph{zero-section.}
We take the existence of a zero-section such that $e(\mathscr{B})\subset\mathscr{Y}$ is Lagrangian as part of our definition of special geometry; for more general situations see \cite{A22}.  Then $\mathscr{Y}$ may be seen as a polarized
commutative group-scheme over the affine scheme $\mathscr{B}$.

The special geometries of interest in this paper
have, in addition, an action $\C\curvearrowright \mathscr{Y}$ by conformal-symplectic automorphisms such that
as $\boldsymbol{\Omega}\mapsto e^{z}\,\boldsymbol{\Omega}$ ($z\in\C$). The quotient group which acts faithfully (and algebraically) is a copy
of the multiplicative group $\C^\times\simeq\C/2\pi i m\Z$. 
The Lie algebra of $\C^\times$ is generated by a Euler vector field $\ce$ such that
\be
\mathscr{L}_\ce\boldsymbol{\Omega}=\boldsymbol{\Omega}
\ee
to which it corresponds a differential $\boldsymbol{\lambda}=\iota_\ce\boldsymbol{\Omega}$
with $d\boldsymbol{\lambda}=\boldsymbol{\Omega}$. 

The smooth fibers $\mathscr{Y}_b$ are connected commutative group-varieties over $\C$.
The universal cover of all such group-varieties is $\C^n$.
Then, analytically, all smooth fibers $\mathscr{Y}_b$ can be written as $\C^n/\Lambda_b$ for some discrete subgroup $\Lambda_b\subset \C^n$. If $(x^1,\cdots, x^n)$ are affine coordinates in the covering $\C^n$,
the Euler differential $\boldsymbol{\lambda}$  locally takes the Darboux form
\begin{equation}\label{kkkas12zzzu}
\boldsymbol{\lambda}=\sum_i p_i\, dx^i\quad\Rightarrow\quad\boldsymbol{\Omega}=dp_i\wedge dx^i
\end{equation} 
for suitable local functions $p_i$ in $\mathscr{B}$ with $\mathscr{L}_\ce\mspace{1mu} p_i=p_i$.

We take the above statements as our basic assumptions (``axioms'').

\paragraph{\bf Physical justifications.} 
All the statements above may be justified (or at least argued)
from the very first principles of quantum physics. In particular the 
equality of the momentum map and the affinization map, eq.\eqref{affini}, follows
from the fact that an $\cn=2$ QFT can be twisted into a Topological
Field Theory (TFT) \emph{\'a la} Witten \cite{ttf1,witten,marino}. After the twist, the TFT has the following properties:
\begin{itemize}
\item[(1)] the IR effective theory is exact for topological amplitudes;
\item[(2)] the TFT amplitudes on $\R^3\times S^1$ are independent
of the radius $R$ of the circle. As $R\to\infty$ we can use the 4d IR effective theory
which is the usual Seiberg-Witten description with Coulomb branch $\mathscr{B}$ \cite{wittenM}. In the limit $R\to0$ we can use the 3d IR effective theory which is
a $\sigma$-model with the hyperK\"ahler target $(\mathscr{Y},\boldsymbol{\Omega})$ \cite{gaiotto2}. The results of the two computations should agree,
in particular we must have the same ring of local topological observables $\mathscr{Q}$ in both descriptions.
This is eq.\eqref{affini}.  
\end{itemize} 

The basic idea is that we may twist the $\cn=2$ theory obtained by
promoting the couplings to \emph{spurion} superfields. For instance, we can see the masses
as the result of very weak gauging of the flavor symmetry and topologically twist this weakly gauged theory. As the gauge coupling goes to zero the Abelian  fibers of the integrable system degenerate into semi-Abelian ones in the well-known way.
 
\medskip

Abstract special geometry (as defined above) captures all sectors of the $\cn=2$ model except
for free hypermultiplets which are decoupled from the rest of the theory and hence
do not talk with the vector multiplets which parametrize the special geometry.

\medskip

This paper is dedicated to understanding
the structure of the geometric objects satisfying the ``axioms''.
We start by a review of the \emph{ordinary} Seiberg-Witten geometries.

\subsection{Review: the case of proper fibers}\label{oridinarycase}

In all abstract special geometries a generic fiber
$\mathscr{Y}_b$ is smooth, hence a complex algebraic group with 
 the property that it
has no non-constant regular functions (cf.\! \eqref{affini}). The obvious way to get rid of all non-constant
regular functions is to take the
fiber to be compact ({proper}). In this case the generic fiber $\mathscr{Y}_b$ is an Abelian variety
with polarization $\omega|_{\mathscr{Y}_b}$. In most applications one assumes the polarization to be principal, but the story makes sense (geometrically as well as physically) for polarizations of any degree. When the generic fiber is proper we use the standard notations and terminology: we write $\mathscr{R}$ (resp.\! $\mathscr{C}$)
for $\mathscr{Q}$
(resp.\! for $\mathscr{B}$) which we call the \emph{chiral ring}
(resp.\! \emph{Coulomb branch}); the total space of the geometry will be written $\mathscr{X}$
with symplectic form $\Omega$.  The dimension $r$ of $\mathscr{C}$ is the \emph{rank} of the geometry.

\subsubsection{Superconformal geometries}

As already mentioned, we are particularly interested in special geometries whose automorphism group contains $\C^\times$. They describe $\cn=2$ SCFTs where all mass and relevant deformations are switched off.
The Lie algebra of the automorphism group $\C^\times$ is generated by a complete holomorphic vector field $\ce$ (the \emph{Euler vector}) such that
\begin{equation}
\mathscr{L}_\ce\mspace{1mu} \Omega=\Omega
\end{equation}
 which implies
\begin{equation}
\Omega=d\lambda_\ce\quad \text{where}\quad \lambda_\ce\equiv \iota_\ce\Omega.
\end{equation} 
The Euler differential $\lambda_\ce$ is the Seiberg-Witten (SW) differential
for SCFT geometries.

The $\C^\times$-action on the regular functions induces a $\C^\times$-action on $\mathscr{C}$.
A Hamiltonian $h\in\mathscr{R}$ has \emph{dimension} $\Delta(h)$ iff $\mathscr{L}_\ce h=\Delta(h)\,h$. The identity has dimension $0$,
 all other elements of $h\in\mathscr{R}$ must have dimension 
 \begin{equation}\label{uyqwza}
 \Delta(h)\geq1
 \end{equation}
 this fact is known as 
 the \emph{unitary bound} (see below).  $\mathscr{R}$ is then a graded ring of the form
\begin{equation}
\begin{aligned}
 &\mathscr{R}=\C\cdot 1\oplus \mathscr{R}_+,&\qquad&\mathscr{R}_+=\bigoplus_{1\leq\Delta\in \frac{1}{n}\mathbb{N}}\mathscr{R}_\Delta,\\ 
 &\mathscr{R}_\Delta\cdot\mathscr{R}_{\Delta^\prime}\subset \mathscr{R}_{\Delta+\Delta^\prime} && h\in\mathscr{R}_\Delta\ \Leftrightarrow\ \mathscr{L}_\ce\,h=\Delta\,h,
 \end{aligned}
\end{equation}
where we used that $n\,\Delta\in\mathbb{N}$ for some integer $n$ since $\C^\times$ acts algebraically. 
It is known that only finitely many $n$ may appear for a given $r$ \cite{A9}. The maximal ideal $\mathscr{R}_+\subset\mathscr{R}$ corresponds to the closed point $0\in\mathscr{C}$, called the \emph{origin,} which is the only closed $\C^\times$-orbit in $\mathscr{C}$.
 If we require $\mathscr{C}$ to be smooth at $0$, we get that $\mathscr{R}\simeq\C[u_1,\dots,u_r]$
 is a free polynomial ring, hence $\mathscr{C}$ is $\C^r$ with coordinates $u_i$
 of dimension $\Delta_i\geq1$. However $\mathscr{C}$ may be singular (cf. \textbf{Question 1}).
 We write $\partial_{\varphi^i}$ for the Hamiltonian vector $v_{u_i}$. The dual differentials $d\varphi^i$
have dimension 
\be
\Delta(d\varphi_i)=-\Delta(v_{u_i})=1-\Delta_i.
\ee
Let $\mathscr{X}_u$ be a smooth fiber. Consider the exponential map
\begin{equation}
 \exp_{e(u)}\mspace{-5mu}\Big(z^i \partial_{\varphi^i}\Big)\colon \mathsf{Lie}(\mathscr{X}_u)\equiv T_{e(u)}\mathscr{X}_u\longrightarrow\mathscr{X}_u,\qquad u\in\mathscr{C}.
\end{equation}
 and let $K_u$ be its kernel. Clearly
\begin{equation}\label{uyqawez}
 \mathscr{X}_u\simeq \mathsf{Lie}(\mathscr{X}_u)/K_u,\quad\text{that is,}\quad \varphi^i\sim \varphi^i+ A(u)^{ij}(m_j+\tau(u)_{jk}\,n^k),\quad m^j,n^k\in\Z,
 \end{equation}
 where $\tau(u)_{ij}$ is the period matrix of the Abelian fiber at $u$, and 
 \be
 \mathscr{L}_\ce\mspace{2mu}\tau_{ij}=0
 \ee
 since $\ce$ acts by automorphisms. The formula \eqref{uyqawez} holds also when $\mathscr{X}_u$ is not smooth,
 except that $\mathsf{Lie}(\mathscr{X}_u)/K_u$ should be identified with the connected component of the smooth locus of the fiber containing $e(u)$.
 Now, 
\begin{equation}\label{pppppeeer}
 \mathscr{L}_\ce\, A^{ij} =(1-\Delta_i)A^{ij}\quad \text{for all }j.
\end{equation}

 \paragraph{Unitary bound.} Let us show eq.\eqref{uyqwza}. All complete vector $v$ satisfies $\mathscr{L}_\ce v =\Delta(v)\,v$ with $\Delta(v)\geq0$, while
 for the Hamiltonian vector $v_h$ such that $\iota_{v_h}\Omega=dh\not\equiv0$ one has
 \begin{equation}
 0\leq\Delta(v_h)\equiv\Delta(h)-1.
 \end{equation} 
The elements $u_i\in\mathscr{R}$ which saturate the unitary bound \eqref{uyqwza}, i.e.\! such that $\Delta_i=1$, have special properties. From eq.\eqref{pppppeeer} we see that $\Delta_i=1$ implies that $A^{ij}$ is constant in the closure of any $\C^\times$-orbit,
 hence constant in $\mathscr{C}$ because $0$ belongs to the closure of all orbits. Since the connected component of the smooth locus of the fiber over $0$
must be a group, it follows that
the fiber over $0$ contains an Abelian variety of dimension equal to the multiplicity of $1$ as a Coulomb dimension, which then is a constant Abelian subvariety $B$ contained in
 all fibers. This Abelian subvariety, describes a free sub-sector which we may decouple without loss. 
 Conversely a constant Abelian subvariety $B$ contained in all the fibers describes a free sector and $\dim B=\dim\mathscr{R}_1$.
 Indeed let $A_\alpha$, $B^\beta$ be a symplectic basis of $H_1(A_0,\Z)$. The periods
 \be
 a^\alpha=\int_{A_\alpha} \iota_\ce\Omega,\qquad a^D_\beta=\int_{B^\beta}\iota_\ce \Omega
 \ee
 are global regular functions on $\mathscr{C}$ of dimension 1.\footnote{\ Notice that the free subsector has no non-trivial mass or relevant deformations.}
  
 We conclude that we may assume with no loss that the chiral ring has
 the form $\C[u_1,\dots,u_r]$, where $r\geq1$ and $\Delta_i>1$ for all $i$.
 In this situation there is a non-zero divisor $\mathscr{D}\subset \mathscr{C}$, called the \emph{discriminant,} consisting of the points $u\in\mathscr{C}$ whose fiber $\mathscr{X}_u$ is not smooth. The complement $\mathring{\mathscr{C}}\equiv\mathscr{C}\setminus\mathscr{D}$ (the ``good'' locus)
parametrizes a family of Abelian varieties, hence over $\mathring{\mathscr{C}}$ we have a local system with fiber $H^1(\mathscr{X}_u,\Z)$.
 The polarization induces a non-degenerate skew-symmetric pairing of this local system
\begin{equation}
 \langle -,-\rangle\colon H_1(\mathscr{X}_u,\Z)\times H_1(\mathscr{X}_u,\Z)\to \Z,\qquad \langle \xi,\eta\rangle=-\langle\eta,\xi\rangle,
\end{equation}
 which is physically interpreted as the Dirac electro-magnetic pairing. The 
 fiber $H_1(\mathscr{X}_u,\Z)$ is the lattice of quantized electro-magnetic charges
 in the SUSY preserving vacuum $u\in \mathring{\mathscr{C}}$. The map
\begin{equation}\label{776zz}
 Z_u\colon H_1(\mathscr{X}_u,\Z)\to \C,\qquad Z_u(\alpha)=\int_\alpha \iota_\ce\Omega,
\end{equation} 
 is the SUSY central charge of a state with electro-magnetic charge $\alpha$.
 
 \subsubsection{The Chow $\mathbb{K}/\C$-trace}
 For later reference we rephrase the saturation of the unitary bound in terms of the unerlying Abelian variety $X_\mathbb{K}$ defined over the function field $\mathbb{K}$. 
 An invariant of an Abelian variety $X_\mathbb{K}$  defined over a complex function field $\mathbb{K}\equiv\C(\mathscr{C})$ is its \emph{Chow $\mathbb{K}/\C$-trace} \cite{trace1,trace2,langII,langIII}
 \be
 \mathrm{tr}_{\mathbb{K}/\C}(X_\mathbb{K})\equiv(B_\C,\tau)
 \ee
 which is an Abelian variety $B_\C$ defined
over $\C$ together with a map defined over $\mathbb{K}$
\be
\tau\colon B_\C\to X_\mathbb{K}
\ee 
which satisfies the appropriate universal mapping property. 
Here we are interested in the physical meaning of the Chow trace (see also \cite{A11,A23}).
We claim that the Chow trace of $X_\mathbb{K}$ is the Abelian variety $B_\C$ of the free subsector.
Indeed under the isomorphism
\be
\Omega\colon \mathsf{Lie}(\mathscr{X}_u)\to T_u^*\mathscr{C}\simeq d\mathscr{R}/d\mathscr{R}^2,
\ee
which sends Hamiltonian vector fields to the differentials of their Hamiltonians,
the image of the Lie algebra of the Chow trace $\Omega(\mathsf{Lie}(B_\C))$
is contained in $d\mathscr{R}_1$ since the Hamiltonian vectors $\partial_w$ tangent to
the Chow-trace have dimension zero. 
Dually all Hamiltonians in $\mathscr{R}_1$ generate constant vertical vector field. 
Thus

\begin{fact} In absence of free subsectors
\be
\mathrm{tr}_{\mathbb{K}/\C}(X_\mathbb{K})=0.
\ee 
\end{fact}

 \subsubsection{Switching on mass and relevant deformations}
 
After switching on the mass and relevant deformations the
 geometry $\mathscr{X}$ remains a holomorphic Lagrangian fibration with section
 over the 
 (same) Coulomb branch $\mathscr{C}$ with generic Abelian fibers. However the $\C^\times$
 symmetry is no longer present, and the Euler holomorphic differential $\lambda_\ce\equiv\iota_\ce\Omega$
 gets replaced by a \emph{meromorphic} Seiberg-Witten differential $\lambda_\text{SW}$
 which depends on the Coulomb branch coordinates $u_i$, the masses $m_a$,
 and the relevant couplings $t_j$.  Eq.\eqref{776zz} still holds with the replacement of $\lambda_\ce$ by $\lambda_\text{SW}$. 
 $\lambda_\text{SW}$ is defined modulo exact forms and 
its dependence on the various parameters follows the rule \eqref{juyuqww123}. 
This rule shows that the tangent space to $\mathscr{C}$ (resp.\! to the space of
 relevant couplings) injects in $H^0(\mathscr{X}_u,\Omega^1)$ (resp.\! $H^1(\mathscr{X}_u,\mathcal{O})$). In particular (as it is clear from the physical side) the space of relevant deformations of a SCFT
 has dimension at most $r$ and is uniquely determined by the $\C^\times$-invariant special geometry of the SCFT. The story with the mass deformations is much more involved. This reflects the physical fact that the mass deformation of a SCFT geometry is non-unique, in general,
 that is, there are several distinct flavor symmetries (of different ranks) which are consistent with one and the same SCFT special geometry.
 This non-uniqueness can be seen \emph{per tabulas} in the explicit classification of the flavor symmetries
 of rank-1 SCFTs in refs.\cite{A3,A4,A5,A6,A7,A8,A11,A13}.

 \medskip
 
 This concludes our quick review of the ordinary special geometries with compact generic fibers (for more see e.g.\! \cite{A23}). Next we consider the general case where the \emph{generic} fiber may be non-compact. 
 
 \section{Geometries with anti-affine fibers}
 
 The ordinary geometries described above are not the only ones which satisfy our physically motivated ``axioms''. The basic condition on the algebraic integrable system that the Liouville
 momentum map is the affinization morphism $\pi$ has other solutions with group-variety fibers.
 One of the goals of this paper is to present the physical interpretation of these more general geometries.
 We first describe them geometrically, beginning with the structure of a \emph{single} generic fiber.


 \subsection{Anti-affine groups}
 An algebraic group is \emph{anti-affine} \cite{milne} iff it has no non-constant regular functions; anti-affine groups are also known as \emph{quasi-Abelian varieties}.
 An anti-affine group is automatically smooth, connected, and commutative \cite{milne}. Our basic ``axiom'',
 eq.\eqref{affini}, says that a general smooth fiber
 $\mathscr{Y}_b$ is an anti-affine group. This is consistent with the interpretation of the 
 geometry as a complex integrable system: the Lie algebra of the fiber is commutative being generated by Hamiltonians in involution.
 However the fibers of most algebraic integrable systems are not anti-affine.
 
 \begin{rem} In the literature there are two distinct notions of ``quasi-Abelian variety''.
 A complex variety is quasi-Abelian in the algebraic sense iff it has no non-constant
 \emph{regular} function. It is quasi-Abelian in the analytic sense iff its underlying complex
 manifold has no non-constant
 \emph{holomorphic} function. Clearly the analytic notion is stronger, and (as we shall see below)
 there exist
 group-varieties which are quasi-Abelian in the algebraic sense but have non-trivial holomorphic functions, in facts whose underlying complex manifold is Stein. We stress that the notion relevant for our applications is the \emph{algebraic one.} 
 The existence of non-trivial holomorphic functions \emph{which are not algebraic} does not spoil the quantum consistency of the QFT associated with the geometry.
 \end{rem}

 To describe the structure of an anti-affine group over $\C$ we start from the structure of
 a general connected, commutative, complex algebraic group $G$ which may or may not be anti-affine. The Barsotti-Chevalley theorem \cite{milne}
 states that $G$ is an extension of group-varieties of the form
\begin{equation}\label{xxxxrrt}
 e\to C\xrightarrow{\phantom{mm}} G\xrightarrow{\sf \;\,al\;} A\to 0,
\end{equation}
 where $A$ is an \emph{Abelian variety,} in facts the \emph{Albanese variety} of $G$ ($\mathsf{al}$ is the Albanese map \cite{langI,milne2}),
 and $C$ is a connected commutative \emph{affine} group with identity $e$.

 A connected affine commutative group $C$ is a product of copies of additive and multiplicative groups
 $\mathbb{G}_a^k\times\mathbb{G}_m^f$; when working over $\C$ we identify it with
 the complex Lie group 
 \begin{equation}
 \C^k\times (\C^\times)^f \equiv V\times T,\quad\text{where}\ \ V=\C^k,\quad T=(\C^\times)^f.
 \end{equation}
We call $T\equiv\mathbb{G}_m^f$ the \emph{torus group} and $V\equiv\mathbb{G}_a^k$ the \emph{vector group}. 
 The inequivalent commutative group-varieties $G$ with given $A$, $k$ and $f$
 are then classified by the group
\begin{equation}
 \mathsf{Ext}^1(A,V\times T)=\mathsf{Ext}^1(A,V)\oplus 
 \mathsf{Ext}^1(A,T).
\end{equation}
 The extension groups are computed in \cite{serre2}
\begin{align}\label{thisone0}
  \mathsf{Ext}^1(A,T)&\simeq \mathsf{Pic}^0(A)^f\equiv (A^\vee)^f\\
   \mathsf{Ext}^1(A,V)&\simeq V^{\oplus\dim A},\label{thisone}
\end{align}
where $A^\vee$ is the dual Abelian variety.
 For later reference we sketch the proof of \eqref{thisone0},\eqref{thisone}.
 
 \begin{proof}[Sketch of proof of \eqref{thisone0},\eqref{thisone}]
 Recall that $\mathsf{Pic}^0(A)$ is the group of divisors on $A$ which are algebraically equivalent to zero modulo linear equivalence,
 while $A^\vee\cong  \mathsf{Pic}^0(A)$ is the \emph{dual} Abelian variety of $A$ (see e.g.\! \S.\,I.8 of \cite{milne2}).
 One shows\footnote{\ See \textsc{Proposition 8.4} of \cite{milne2}, or \S.\,5.2 of \cite{langII}.} that an invertible sheaf ($\equiv$ line bundle $\equiv$ linear class of divisors) $\cl\in \mathsf{Pic}(A)$ is algebraically trivial if and only if it is invariant under translations in
 $A$, that is, if for all $a\in A(\C)$
 we have
 \begin{equation}\label{kkkkasaz12ii}
 t_a^*\cl \simeq \cl,\quad\text{where }t_a\colon b\mapsto b+a,\ \ a,b\in A(\C).
 \end{equation} 
 We first consider extensions of $A$ by $\mathbb{G}_m$. Complex analytically, $G$ is a principal bundle over $A$
 with structure group $\C^\times$, which we may see as a bundle $\cl^*\to A$ where $\cl^*$ is the complement of the zero section
 in a line bundle $\cl\to A$. The total space of a principal bundle over an Abelian group $A$ with Abelian structure group $\C^\times$
 is itself an Abelian group if and only if the line bundle is invariant by translation on $A$ so that the two group operations
  along the base and  the fiber
 commute; by the remark before eq.\eqref{kkkkasaz12ii}
  this is equivalent to $\cl$ being algebraically trivial.
Next we consider extensions of $A$ by $\mathbb{G}_a$. Complex analytically $G$ is a principal bundle over $A$ with group $\C$, and all such principal bundles are commutative groups. Hence
the extension group is the space of isomorphism classes of such bundles $H^1(A,\co_A)\simeq \C^{\dim A}$.
 \end{proof}
 
 We note that
 \begin{equation}
 H_1(G,\Z)\simeq \Z^{2\dim A+\dim T}.
 \end{equation}

The algebraic group $G$ in eq.\eqref{xxxxrrt} is connected and commutative but may or may not be anti-affine. 
 For instance if $A$ is trivial, $G$ is affine; more generally when the extension is trivial, $G=C\times A$,
 the group is not anti-affine, etc. Our next task is to understand which extensions are anti-affine.
We consider first the extension of $A$ by the $f$-torus $T\equiv (\C^\times)^f$ and then the extension by a vector group $V\simeq \C^k$.  
  
  \subsubsection{Torus extensions of Abelian varieties}
  If $G$ is an extension of the Abelian variety $A$ by the torus group $T$
  \begin{equation}\label{4443311}
  e\to T\to G\xrightarrow{\sf \;al\;} A\to 0
  \end{equation}
the Albanese map $G\xrightarrow{\sf \;al\;} A$ is a principal $T$-bundle. We ask when a $T$-principal bundle over a complex Abelian variety $A$ is a commutative
algebraic group.
Let $\Lambda$ be the character group of $T\simeq (\C^\times)^f$.
For any $T$-principal bundle $G\xrightarrow{\,\alpha\,} A$ we have the ``Fourier decomposition''
\begin{equation}\label{kkkkkssss23}
\alpha_\ast\mspace{1mu}\co_G =\bigoplus_{\lambda\in \Lambda} \cl_\lambda,
\end{equation}
where $\cl_\lambda\to A$ are line bundles (invertible sheaves).
Around eq.\eqref{kkkkasaz12ii} we saw that the translations in $A$ are
$T$-automorphisms of the principal bundle $G$ iff
the line bundles $\cl_\lambda\to A$ are algebraically equivalent to zero,
 that is, $\cl_\lambda\in \mathsf{Pic}^0(A)$ for all $\lambda\in\Lambda$. 
Stated differently:

\begin{lem}[See e.g.\! \cite{brion}] The $T$-principal bundle $G\xrightarrow{\; \alpha\; }A$ is a group homomorphism with kernel $T$ if and only if the decomposition \eqref{kkkkkssss23} defines a group homomorphism
\begin{equation}\label{lllk90n}
\chi\colon \Lambda\to \mathsf{Pic}^0(A),\qquad \chi\colon \lambda\mapsto \cl_\lambda.
\end{equation}  
\end{lem}

Then we have
\begin{equation}
\Gamma(G,\co_G)= H^0(A,\alpha_\ast\co_G)=\bigoplus_{\lambda\in \Lambda} H^0(A,\cl_\lambda)
\end{equation}
while
\begin{equation}
H^0(A,\cl_0)\equiv H^0(A,\co_A)=\C.
\end{equation} 

Hence $G$ is anti-affine iff $H^0(A,\cl_\lambda)=0$ for all  $\lambda\neq0$. Since 
 $H^0(A,\cl)=0$ for all line bundle $\cl$ which is algebraically trivial but
non-trivial, we need that $\cl_\lambda$ is the zero element of $\mathsf{Pic}^0(A)$ only when
$\lambda=0$, that is,

\begin{lem}\label{uytz04} The group $G$ in eq.\eqref{4443311} is anti-affine if and only if the underlying  group homomorphism
$\chi\colon \Lambda\to \mathsf{Pic}^0(A)$ is injective, i.e.\! if and only if
$\Lambda$ is a lattice in the dual Abelian variety $A^\vee$.
\end{lem}


An example will clarify the situation.


\begin{exe}[Picard 1910]\label{kkkasqw12} We consider the extensions of the elliptic curve $E_\tau$ of period $\tau$
by the one-dimensional torus $\C^\times$. We see $E_\tau$ as the quotient of its universal cover, $\C$,
by the group generated by the two automorphisms 
\begin{equation}
L_1\colon z\mapsto z+1\quad\text{and}\quad L_2\colon x\mapsto z+\tau,
\end{equation}
and we write $\pi\colon \C\to E_\tau\equiv \C/\langle L_1,L_2\rangle$ for the canonical projection.
Let
$p_1\colon\C\times \C\to \C$ be the trivial $\C$-bundle. A line bundle $\cl_w\to E_\tau$ which is algebraically trivial can always be written as the quotient
\begin{equation}
\cl_w= (\C\times \C)\big/\langle L_1,L_2\rangle,\qquad \begin{aligned}L_1&\colon (z,y)\mapsto (z+1,y)\\
L_2&\colon (z,y)\to (z+\tau, e^{2\pi i w}y).
\end{aligned}
\end{equation}
From this expression it is obvious
 that the inequivalent bundles $\cl_w\in\mathsf{Pic}^0(E_\tau)$
 are parametrized by the points  $w\in E_\tau$ (which coincides with $E_\tau^\vee\simeq\mathsf{Pic}^0(E_\tau)$). To get the corresponding $\C^\times$ extension of $E_\tau$ we just cut out the zero section of the bundle 
 \begin{equation}
G_w= (\C\times \C^\times)\big/\langle L_1,L_2\rangle,\qquad \begin{aligned}L_1&\colon (z,y)\mapsto (z+1,y)\\
L_2&\colon (z,y)\to (z+\tau, e^{2\pi i w}y).
\end{aligned}
\end{equation}
It may be convenient to write the principal $\C^\times$-bundle $G_w\to E_\tau$
in an alternative way which we dub the ``singular gauge'' ; we parametrize 
\begin{equation}\label{uytqwaezz}
(z,y)=\left(z, \frac{\vartheta(z-w,\tau)}{\vartheta(z,\tau)} x\right)\in \C\times \big(\mathbb{P}^1\setminus\{0,\infty\}\big)
\end{equation}
where\footnote{\ The $\theta$-function in the \textsc{rhs} is defined as in chapter 20 of DLMF \cite{dlmf}.
By construction $\vartheta(z,\tau)$ has a single zero at the lattice points $z=m+n\tau$.}
\begin{equation}
\vartheta(z,\tau)\overset{\rm def}{=} \theta_3\big(\pi z-\tfrac{1}{2}\pi -\tfrac{1}{2}\pi\tau\,\big|\, \tau\big)
\end{equation}
and then take the quotient by the action of the group $\langle \tilde L_1,\tilde L_2\rangle$ acting as
\begin{equation}
\tilde L_1\colon (z,x)\mapsto (z+1,x),\qquad \tilde L_2\colon (z,x)\mapsto (z+\tau,x),
\end{equation}
where $x\in\C^\times$ is now a global (but singular) coordinate along the fiber. 

It is clear that the groups extensions $G_w$
are parametrized by degree-zero divisors of the form
\begin{equation}
\left(\frac{\vartheta(z-w,\tau)}{\vartheta(z,\tau)}\right)=w-0,
\end{equation}
i.e.\! by points of $E_\tau^\vee\cong E_\tau$.
We conclude:
\begin{corl} Modulo isomorphism we have \emph{one} connected, commutative,
 algebraic group (over $\C$), $G_w$, which fits in the exact sequence
\begin{equation}\label{kasqwertooo8}
1\to \C^\times \to G_w\to E_\tau\to 0
\end{equation}
per point $w\in E_\tau\cong \mathsf{Pic}^0(E_\tau)$.
\end{corl}
The \textbf{Corollary} is just eq.\eqref{thisone0} with $f=1$.
Now let us see for which points $w\in E_\tau$ the commutative
 algebraic group $G_w$ is \emph{anti-affine.} Suppose that the point $w\in E_\tau$ which defines the group $G_w$ is \emph{torsion,} that is,
$m w=0$ in $E_\tau$ for some $m\in\Z$, i.e. 
\begin{equation}
m\, w= a+b\,\tau\ \text{in }\  \C\ \text{with $a,b\in\Z$.}
\end{equation} 
The function on $\C\times \C^\times$
\begin{equation}
f(z,y)\equiv e^{-2\pi i b z}\,y^m
\end{equation} is invariant under $L_1$, $L_2$ so $f(z,y)$ is a non-constant global holomorphic function and $G_w$ is \emph{not}
anti-affine. However, when $w$ is not torsion $G_w$ is manifestly anti-affine.\footnote{\ It is also quasi-Abelian in the analytic sense.} This
conclusion is in agreement with \textbf{Lemma \ref{uytz04}} (see \cite{brion}
for more details).

The dual of the Lie algebra of the 2-dimensional group $G_w$ is generated by two 
holomorphic differentials $dz$ and $dy/2\pi i y$. $H_1(G_w,\Z)\simeq \Z^3$ is generated by the three  cycles
\begin{equation}
\begin{aligned}
A&=\{(z(s),y(s))=(s,1)\}, && B=\{(z(s),y(s))=(\tau s, e^{2\pi i s w})\},\\
C&=\{(z(s),y(s))=(0, e^{2\pi is})\},&&\text{where}\ \ 0\leq s\leq 1
\end{aligned}
\end{equation}
with period matrix
\begin{equation}\label{perper}
\begin{aligned}
\int_A dz&=1, &\int_B dz&=\tau, &\int_C dz&=0,\\
\int_A \frac{dy}{2\pi i y}&=0, &\int_B \frac{dy}{2\pi iy}&=w, &\int_C \frac{dy}{2\pi i y}&=1,
\end{aligned}
\end{equation}  
and the groups $G_w$ are distinguished by their period of $dy/2\pi i y$ on the $B$-cycle.
$G_w$ is anti-affine (quasi-Abelian) when this period cannot be written as $a +b\tau$ for $a,b\in\mathbb{Q}$, i.e.\! when the three non-zero periods \eqref{perper} are linearly independent over $\mathbb{Q}$.
\end{exe}

Torus extensions of Abelian varieties are also called \emph{semi-}Abelian varieties.
They may be seen as semistable degenerations of Abelian varieites.

\subsubsection{Vector extensions of Abelian varieties}

For $A$ an Abelian variety over $\C$ we have
\begin{equation}
\mathsf{Ext}^1\mspace{-2mu}(A,\C)\simeq \mathsf{Lie}(A),
\end{equation}
so that an extension of $A$ by a vector space $V$ of dimension $k$ is specified by a $k$-tuple of elements of the 
Lie algebra of $A$. More precisely

\begin{pro}[see \cite{brion} \textbf{Proposition 2.3}]\label{vectex} All extension $G$ of an Abelian variety $A$
by a vector group $V$ fits in a unique commutative diagram
\begin{equation}
\begin{gathered}
\xymatrix{0\ar[r] & H^1(A,\co_A)^\vee \ar[d]_\gamma \ar[r] & E(A)\ar[d] \ar[r] & A\ar@{=}[d]\ar[r] &0\\
0 \ar[r]& V \ar[r] & G \ar[r] & A\ar[r] &0}
\end{gathered}
\end{equation}
where $E(A)$ is the \emph{universal vector extension of $A$ \cite{ros58,MM}} (which is an anti-affine group).
Then $G$ is anti-affine if and only if $\gamma$ is surjective. The transpose map
\begin{equation}
\gamma^t\colon V^\vee\to H^1(A,\co_A)
\end{equation}
is then injective and we conclude that the anti-affine groups over $A$
obtained as vector extensions are classified by subspaces of the $\C$-space 
\begin{equation}
H^1(A,\co_A)\simeq H^0(A,\Omega^1_A)^\vee\simeq (T^*_eA)^\vee\equiv T_eA \simeq \mathsf{Lie}(A)
\end{equation}
where the first isomorphism is given by the polarization, the second one by translation invariance
in the Abelian group $A$.
\end{pro}

This statement is crucial for the physical interpretation of vector extensions in the context of $\cn=2$ QFT.
We have non-trivial group extensions of an Abelian variety $A$ by any vector space $\C^k$
of arbitrary large dimension $k$ \emph{but} if we insist that the extension group should be
\emph{anti-affine,} we conclude that $k$ cannot be larger than $r\equiv\dim A$.

\begin{rem} The vector extensions are quasi-Abelian varieties in the algebraic
sense but not in the analytic sense. In other words, while they have no non-constant
\emph{regular} functions they have plenty of non-constant \emph{holomorphic} functions, indeed the underlying complex manifold of the universal vector extension is Stein see the \textbf{Example} below. 

\end{rem}

\begin{exe}\label{addtive}
We consider the additive version of the  Picard construction for the group
\begin{equation}
\xymatrix{0\ar[r] & \C \ar[r] & H_\eta\ar[r] & A\ar[r] & 0}
\end{equation}
\begin{gather}
H_\eta=( \C\times \C)/\langle L_1,L_2\rangle\intertext{where}
L_1\colon (z,y)\to (z+\omega_1,y+\xi_1),\qquad L_2\colon (z,y)\to (z+\omega_2, y+\xi_2).
\end{gather}
$H_1(H_\eta,\Z)$ is generated by the two cycles $A=\{s\,\omega_1,s\,\xi_1\}$ and $B=\{s\,\omega_2,s\,\xi_2\}$.
We define
\begin{equation}\label{hqwert}
\eta \equiv \frac{1}{2\pi i}\left(\int_A dz\int_B dy-\int_A dy\int_B dz\right)=\frac{\omega_1\,\xi_2-\omega_2\,\xi_1}{2\pi i},
\end{equation}
Again, formulae look simpler in a ``singular gauge''. We write
\begin{equation}\label{kkkkhfii9}
(z,y)=\big(z, x+\eta\, \zeta(z)\big)
\end{equation}
subject to the identifications
$(z,x)\sim (z+\omega_1,x)\sim (z+\omega_2,x)$.
In eq.\eqref{kkkkhfii9} $\zeta(z)$ is the Weierstrass $\zeta$-function which satisfies
\begin{equation}\label{weier}
\frac{d}{dz} \zeta(z)= -\wp(z)
\end{equation}
with $\wp(z)$ the Weierstrass function on $E_\tau$ of periods $\omega_1$, $\omega_2$
and $\tau=\omega_2/\omega_1$.
The condition that $H_\eta$ is anti-affine in the algebraic sense is
$\eta\neq0$: indeed in this case 
no polynomial in $y$ with coefficients holomorphic in $z$ is invariant under $L_1$, $L_2$. However 
we have a ring of (non-algebraic) non-trivial global holomorphic functions
\begin{equation}
\psi_{m,n}(z\,y)=\exp\!\left[\frac{m(\xi_1 z-\omega_1y)+n(\xi_2 z-\omega_2y)}{\eta}\right],\quad
m,n\in\Z.
\end{equation}
One checks that these functions separate points in 
$H_\eta$: indeed as a complex manifold $H_\eta$ is known to be Stein.

From eq.\eqref{hqwert} it is clear that $\eta$ measures the projection in $H^{0,1}(E)$
of the cohomology class of the second-kind differential $dy$, in agreement with \textbf{Proposition \ref{vectex}}
which says that the extension by $\C$ of $E$ are classified by $H^1(E,\co_E)\simeq H^{0,1}(E)$.  
\end{exe}

\subsection{General anti-affine Lagrangian fibrations}

Our next task is to study Lagrangian fibration with general (algebraic) quasi-Abelian fibers.

We first consider a fibration 
\be\label{ourfibration}
\pi\colon \mathscr{Y}\to \mathscr{B}
\ee
 whose
general fiber $\mathscr{Y}_b$ is a general (algebraic) connected commutative
complex group, not necessarily anti-affine, which is a Lagrangian submanifold, $\boldsymbol{\Omega}|_{\mathscr{Y}_b}=0$. As in \S.\,\ref{oridinarycase}  in addition
 we assume the existence of
a complete Euler vector $\ce$ which generates a group $\C^\times$ of
automorphisms of all the relevant structures with $\mathscr{L}_\ce\, \boldsymbol{\Omega}=\boldsymbol{\Omega}$.
As before this leads to an action on the zero-section $\mathscr{B}$ and, just as in \S.\,\ref{oridinarycase},
when $\mathscr{B}$ is smooth it is the spectrum of a polynomial ring $\mathscr{Q}=\C[x_1,\dots,x_\ell]$
with $\ell=\dim \mathscr{Y}_b$. So $\mathscr{B}=\mathsf{Spec}\,\mathscr{Q}$
and $\mathscr{L}_\ce\, x_s= \Delta(x_s)\,x_s$ with $\Delta(x_s)>0$ for all $s$.

By the Barsotti-Chevalley theorem, 
the generic fiber $\mathscr{Y}_b$ is the extension of
an Abelian variety $\mathscr{A}_b$ of dimension $r$ by a vector group $V$ of dimension $k$
and a torus group $T$ of dimension $f$ ($\ell=r+k+f$). The Lie algebra of the fiber has the form
\begin{equation}\label{lieeeea}
\mathsf{Lie}(\mathscr{Y}_b)\equiv T_{e(b)}\mathscr{Y}_b= \mathsf{Lie}(\mathscr{A}_b)\oplus V\oplus \mathsf{Lie}(T),
\end{equation}
while the symplectic form yields an inverse pair of isomorphisms
\be
\xymatrix{T_{e(b)}\mathscr{Y}_b \ar@/^1pc/[rr]^{\boldsymbol{\Omega}} && \ar@/^1pc/[ll]^{\boldsymbol{\mho}} T_b^*\mathscr{B}\ar@{=}[r]&\bigoplus_s \C\, dx_s}
\ee
which define three Pfaff systems $\boldsymbol{\Omega}\, \mathsf{Lie}(A_s)$,
$\boldsymbol{\Omega}\, V$ and $\boldsymbol{\Omega}\, \mathsf{Lie}(T)$ in $\Omega^1(\mathscr{B})$
which are integrable since the fibers' Lie algebra is commutative. Then 
we have 
\begin{equation}
(x_1,\dots,x_\ell)=(t_1,\dots,t_k,m_1,\dots,m_f,u_1,\dots,u_r)
\end{equation}
with
\begin{equation}
\boldsymbol{\Omega}\, V=\bigoplus \C\, dt_j,\quad \boldsymbol{\Omega}\, \mathsf{Lie}(T)=\bigoplus \C\, dm_a
\quad \boldsymbol{\Omega}\, \mathsf{Lie}(A_s)=\bigoplus \C\, du_i.
\end{equation}
and the symplectic form may be written as
\begin{equation}\label{iiiiiiqqqw}
\boldsymbol{\Omega}= d\alpha^j \wedge dt_j+ \frac{dy^a}{2\pi i y^a}\wedge dm_a+d z^i\wedge du_i,
\end{equation}
where $\alpha^j$, $y^a$ and $z^i$ are coordinates along the orbits of the dual Hamiltonian vector fields.   
As before we may assume $\Delta(u_i)>1$ without loss. This is all we can say without further restrictions on the commutative groups which are allowed as fibers. However, as we argued, physics requires in addition
the fibers to be \emph{anti-affine.} 

\subsubsection{Vector extension vs.\! relevant interactions}
We  consider the vector extension first and then the torus one.
We write $\partial_{z^i}$ for the vector fields $\mho\, du_i$ which generate the Lie algebra of $\mathscr{A}_b$. One has
\begin{equation}
\Delta(\partial_{z^i})=\Delta(u_i)-1.
\end{equation}
 $H^0(\mathscr{A}_b,\Omega^1)$ is spanned by the dual differentials $dz^i$ with
\begin{equation}
\Delta(dz^i)=1-\Delta(u_i).
\end{equation}
By \textbf{Proposition \ref{vectex}} the summand $V\equiv\bigoplus\C \,\partial_{\alpha_j}$ in the \textsc{rhs} of \eqref{lieeeea} 
is a quotient of
\be H^1(A,\co_A)^\vee\simeq \bigoplus_i\C\, dz^i.
\ee
 Let $dz^{i(j)}$ be
 a representative of $\partial_{\alpha_j}$; we have
\begin{equation}
\Delta(\partial_{\alpha_j})= \Delta(d z^{i(j)})=1-\Delta(u_{i(j)})
\end{equation}
and the dimension of the coordinate $t_j$ such that $dt_j=\Omega\, \partial/\partial\alpha_j$ is
\begin{equation}
\Delta(t_j)= 1+\Delta(\partial_{\alpha_j})=2-\Delta(u_{i(j)})
\end{equation}
that is,
\begin{equation}\label{predphys}
\Delta(t_j)+\Delta(u_{i(j)})=2.
\end{equation}
Thus the condition that the fiber is anti-affine associates to each summand $\C \,\partial_{\alpha_j}$ in $V$
a distinct operator $u_{i(j)}$ with the dimension \eqref{predphys} predicted by physics (this conclusion will be confirmed below by the explicit construction of the SW differential).
Since the ring $\mathscr{Q}$ is positively graded, $\Delta(t_j)>0$ and we see that 
the additive extension is at most given by one new extra coordinate $t_j$
(coupling) per chiral operator $u_i$ of the undeformed SCFT with dimension 
\be
1< \Delta(u_i) <2.
\ee
 A chiral operator $\phi\in\mathscr{R}$ 
with dimension in this range is automatically a generator of the polynomial ring,\footnote{\ Under our standing assumption that no element of $\mathscr{R}$ has dimension 1.} i.e.\! one of the $u_i$'s. 
We stress that this correspondence between the deformation parameters $t_j$
and ``relevant operators'' in $\mathscr{R}$ follows from the requirement
that the fiber is anti-affine. Without the condition of anti-affinity,
no relation between chiral operators and coupling parameters would exist and we would get physically meaningless geometries.

\begin{rem} The conclusion holds whether the Coulomb branch is smooth or not.
\end{rem}

\subsubsection{Torus extensions vs. flavor symmetry}

Let us now turn to the subtler torus extensions. The Lie algebra of $\C^\times$ is generated by $y\,\partial_y$,
so the Hamiltonian vector fields  naturally have dimension zero. The corresponding Hamiltonians $m_a$, i.e.\! coordinates on the base $\mathscr{B}$, then have $\Delta(m_a)=1$, which is consistent
with their interpretation as mass parameters. The
physical masses take value in the complexified dual of the Cartan subalgebra of the
flavor symmetry group $F$. In particular $f\equiv\dim T$ should coincide with the rank of the flavor group $F$.

The anti-affine condition requires the character group of the torus $T$
to embed as a lattice $\chi(T)$ in the Picard variety $\mathscr{A}_b^\vee$ for general $b\in\mathscr{B}$. We wish to interpret the lattice $\chi(T)$
as the weight/root lattice of
the flavor symmetry $F$. Just as for the relevant deformations, we expect that in each given rank $r$ ($\equiv$ the dimension of the Albanese variety of the general fiber)
 there is
a maximal dimension $f$ of the extending torus $T$ which is consistent with the fiber being an anti-affine algebraic group.
In facts, we expect that a $\C^\times$-isoinvariant
 special geometry in the standard sense describes finitely many
SCFTs, each one with its own flavor symmetry, so that only finitely many maximal torus extensions are allowed for a given ordinary special geometry, and hence there if a maximal dimension $f_\text{max}$ which depends on
the invariants of the ordinary special geometry to be deformed. To make this idea concrete
we use the Mordell-Weil group \cite{MW1}, generalizing to arbitrary $r$ the discussion in ref.\cite{A11}
for $r=1$. We shall return to the flavor issue after
reviewing the Mordell-Weil lattice.

\section{The Mordell-Weil lattice}

Our fibration \eqref{ourfibration} is a model of an anti-affine group variety $Y_\mathbb{L}$ defined over the function field $\mathbb{L}\equiv\C(\mathscr{B})$.
Its Albanese variety $A_\mathbb{L}$ has a model over $\C$, $\mathscr{A}\to\mathscr{B}$,
which is the universal special geometry of the $\cn=2$ QFT parametrized by the space $\mathscr{P}$ of couplings $t_j,m_a$, i.e.
\be
\begin{gathered}
\xymatrix{\mathscr{A}\ar[rr]\ar[dr]&& \mathscr{B}\simeq \mathscr{C}\times \mathscr{P}\ar@/^1pc/[ld]\ar[r] &\mathscr{C}\\
& \mathscr{P}}
\end{gathered}\label{fiffirr}
\ee
 $A_\mathbb{L}$ is polarized, and as customary we assume the polarization to be principal for simplicity. This assumption is not important. Anyhow it allows us to identify $A_\mathbb{L}$
and its dual $A_\mathbb{L}^\vee$. $Y_\mathbb{L}$ is the extension of $A_\mathbb{L}$
by the affine commutative group $\mathbb{G}_m^f\times \mathbb{G}_a^k$. We focus on the multiplicative part of the story since the vector extensions are already fully understood, so set $k=0$.
Since $Y_\mathbb{L}$ is anti-affine we have a group embedding, defined over $\mathbb{L}$,
\be\label{uyqwert}
\chi\colon C\to A_\mathbb{L}^\vee(\mathbb{L})\simeq A_\mathbb{L}(\mathbb{L})
\ee
of the character group $\simeq \Z^f$ of $\mathbb{G}_m^f$ into the group $A_{\mathbb{L}}(\mathbb{L})$ of $\mathbb{L}$-valued
points of $A_\mathbb{L}$.
The image $\chi(C)$ is a rank-$f$ sub-lattice contained
in the \emph{Mordell-Weil lattice} 
\be 
\mathsf{MWL}(A_\mathbb{L})\;\overset{\rm def}{=}\; A_{\mathbb{L}}(\mathbb{L})\big/ A_{\mathbb{L}}(\mathbb{L})_\text{torsion}
\ee
of the Abelian variety $A_\mathbb{L}$.

In the language of the complex model $\varpi\colon\mathscr{A}\to\mathscr{B}$ of $A_\mathbb{L}$, the generic smooth fiber $\mathscr{A}_b$
 is a polarized Abelian variety of complex dimension $r$, and we have
 a rank-$f$ sublattice $\chi(C)\subset\mathscr{A}_b$. 
Varying $b$ in $\mathscr{B}$ each element of $C$ defines a section of $\varpi$.
 The Mordell-Weil group
 is then realized as a group of global sections of $\mathscr{A}\to\mathscr{B}$ \cite{MW1},
 and $\chi(C)\subset \mathscr{A}$ is a group of sections over the base $\mathscr{B}$. 
 
\subsection{The Chow $\mathbb{L}/\C$-trace and the Lang-N\'eron theorem}An important invariant of an Abelian variety $A_\mathbb{L}$  defined over a complex function field $\mathbb{L}\equiv\C(\mathscr{B})$ is its \emph{Chow $\mathbb{L}/\C$-trace} \cite{trace1,trace2,langI,langII,langIII}: $\mathrm{tr}_{\mathbb{L}/\C}(A_\mathbb{L})\equiv(B_\C,\tau)$ is an Abelian variety $B_\C$ defined
over $\C$ together with a map defined over $\mathbb{L}$
\be
\tau\colon B_\C\to A_\mathbb{L}
\ee 
which satisfies the appropriate universal mapping property. 
Here we are interested in the physical meaning of the Chow trace (see also \cite{A23}).
We specialize the geometry 
 to a point $p\in\mathscr{P}$  in parameters' space getting a
a Lagrangian fibration 
\be
\mathscr{X}\equiv\mathscr{A}_p\to\mathscr{C}
\ee
which is the special geometry of the $\cn=2$ QFT with fixed couplings $p$ (see eq.\eqref{fiffirr}).
We focus on the underlying Abelian variety $A_\mathbb{K}$ over the function field $\mathbb{K}\equiv\C(\mathscr{C})$ and
 identify the generic fiber $\mathscr{X}_u$ with $A_\mathbb{K}$.
Then
$\mathscr{X}_u$ contains the Chow $\mathbb{K}/\C$-trace $B_\C$ of $A_\mathbb{K}$
as an Abelian subvariety. The total space of $\mathscr{X}$ is a symplectic variety with symplectic form 
$\Omega$ and smooth fibers complex groups. Under the isomorphism
\be
\Omega\colon \mathsf{Lie}(\mathscr{X}_u)\to T_u^*\mathscr{C}\simeq d\mathscr{R}/d\mathscr{R}^2,
\ee
which sends Hamiltonian vector fields to the differentials of their Hamiltonians,
the image of the Lie algebra of the Chow trace $\Omega(\mathsf{Lie}(B_\C))$
is contained in $d\mathscr{R}_1$ since the Hamiltonian vectors $\partial_w$ tangent to
the Chow-trace have dimension zero: 
\be
\mathscr{L}_\ce\partial_w=0.
\ee
Indeed the image is precisely $d\mathscr{R}_1$. Since the elements of $\mathscr{R}_1$
correspond to the free subsector of the QFT, we conclude that the Chow trace is the fiber
of the free subsector.\footnote{\ Notice that the free subsector has no non-trivial mass or relevant deformations.}  Since we are assuming (with no loss) that our geometry does not
contain free subsectors, i.e.\! $\mathscr{R}_1=0$, the Chow trace is automatically trivial.

When the Chow trace $\mathrm{tr}_{\mathbb{L}/\C}(A_\mathbb{L})$ is trivial, 
 the Lang-N\'eron theorem \cite{NL} guarantees that 
the Mordell-Weil group $\mathsf{MW}(A_\mathbb{L})$ is a finitely generated group. Its free part is the \emph{Mordell-Weil lattice} \cite{MW1} 
\be
\mathsf{MWL}(A_\mathbb{L})=\mathsf{MW}(A_\mathbb{L})/\mathsf{MW}(A_\mathbb{L})_\text{torsion}.
\ee

Comparing with the discussion around eq.\eqref{uyqwert} we conclude:
\begin{fact} An \textbf{anti-affine} extension $Y_\mathbb{L}$ of the Abelian variety $A_\mathbb{L}$
defined on the function field $\mathbb{L}=\C(\mathscr{B})$ with $\mathrm{tr}_{\mathbb{L}/\C}(A_\mathbb{L})=0$ ($\equiv$ no free subsectors) by the torus group $\mathbb{G}_m^f$
corresponds to a rank-$f$ sublattice
\be
\Lambda\subset A(\mathbb{L})/A(\mathbb{L})_\text{tor}\equiv \mathsf{MWL}(A_\mathbb{L})
\ee
 of the Mordell-Weil lattice of
$A_\mathbb{L}$. In particular the rank $f$ of the torus extension is at most equal to $\mathrm{rank}\,\mathsf{MWL}(A_\mathbb{L})$. 
\end{fact}

We see that the anti-affine condition goes in the desired direction of putting an upper bound on the rank of the extension (which we wish to identify with the rank of the flavor group $F$).

\subsubsection{N\'eron-Tate height}

The lattice $\mathsf{MWL}(A_\mathbb{L})$ is endowed with a non-degenerate,
positive-definite, symmetric bilinear pairing called the \emph{N\'eron-Tate height}
(or canonical height) \cite{langII,langIII}.

\begin{thm}[e.g.\! {\bf Theorem 5.4.4} in \cite{langII}]\label{poiuyt} Let $\Gamma$ be a finitely generated subgroup of
\begin{equation}
A_\mathbb{L}(\mathbb{L})\big/(\tau B_\C(\C)+A_\mathbb{L}(\mathbb{L})_\text{\rm torsion}\big).
\end{equation}
The canonical height is non-degenerate on $\Gamma$, and its extension to
$\R\otimes A_\mathbb{L}(\mathbb{L}^{\rm al})/\tau B_\C(\C)$ is a positive definite quadratic form.
($\mathbb{L}^{\rm al}$ is an algebraic closure of $\mathbb{L}$).
\end{thm}

As a corollary the lattice $\Lambda$ is
 endowed with a symmetric positive-definite pairing
\begin{equation}
\langle -,-\rangle\colon \Lambda\times \Lambda\to \Z.
\end{equation}  

The physical intuition says that the rank-$f$ lattice $\Lambda$ equipped with the N\'eron-Tate (NT) quadratic form should
be identified with the lattice of weights of the (visible) flavor Lie group $F$ associated to the mass parameters
in the space $\mathscr{P}$. This was shown to be the case when $r=1$ \cite{A11}.
The weight lattice of $F$ carries a positive-definite quadratic form
(invariant under the action of the Weyl group) which we wish to identify with the N\'eron-Tate form
(possibly up to normalization) \cite{A13}. 

\subsection{The rank of the torus extension}

We wish to say more on the rank of the Mordell-Weil lattice of $A_\mathbb{L}$. In the classical case, i.e.\! in dimension $r=1$, the Mordell-Weil rank is given by the Shioda-Tate formula \cite{MW1}. There is a generalization
of this formula valid in higher dimension
due to Oguiso \cite{Oguiso} (see also \cite{ST2,ST3}).

Th Oguiso theorem applies in the following situation:
we have a fibration (with section) $\pi\colon W\to Z$, with $W,Z$ normal projective, such that
the generic fiber $A_{\C(Z)}\equiv W_\eta$ is an Abelian variety (of positive dimension), defined over $\C(Z)$,
with the properties:
\begin{itemize}
\item[(1)] $W$, $Z$ have only $\mathbb{Q}$-factorial canonical singularities;
\item[(2)] there is no prime divisor $D\subset W$ such that
$\dim \pi(D)\leq \dim Z-2$;
\item[(3)] $h^1(W,\co_W)=h^1(Z,\co_Z)=0$.
\end{itemize}
In our case $Z$ is typically $\mathbb{P}^{\mspace{2mu}r}$ while $W$ is a suitable projective variety
such that $\mathscr{A}=W\setminus E$ for a divisor $E$. The condition (3) is automatically satisfied in absence of
free subsectors (i.e.\! $A_{\C(Z)}$ has zero Chow $\C(Z)/\C$-trace) which is a necessary condition for the Mordell-Weil group $A_{\C(Z)}(\C(Z))$ to be finitely generated. (1),(2) are mild regularity conditions that we assume to hold (recall that in this paper we ignore all singularities as long as they are confined in codimension at least 2).

Let $\cd\subset Y$ be the discriminant locus, $\cd=\sum_i D_i$
and 
\begin{equation}
\pi^*(D_i)_\text{red}=\sum_{j=0}^{m_i-1} D_{ij}
\end{equation}
be the respective decompositions into irreducible components.
$m_i$ is the number of prime divisors on $W$ over $D_i\subset Z$.
Then

\begin{thm}[Refs.\cite{Oguiso,ST2,ST3}] Under the assumptions {\rm(1)-(3)}, the Mordell-Weil group 
$\mathsf{MW}(A_{\C(Z)})$ is a finitely generated Abelian group of rank
\begin{equation}\label{Ofurmula}
\mathrm{rank}\, \mathsf{MW}(A_{\C(Z)})=\rho(W)-\rho(Z)-\mathrm{rank}\, \mathsf{NS}(A_{\C(Z)})-\sum_{i=1}^k(m_i-1)
\end{equation}
Here $\rho(W)$ (resp.\! $\rho(Z)$) is the Picard number of $W$ (resp.\! of $Z$)
and $\mathsf{NS}(A_{\C(Z)})$ is the N\'eron-Severi group of $A_{\C(Z)}$. In particular,
\begin{equation}
\mathsf{rank}\, \mathsf{MW}(A_{\C(Z)})\leq \rho(W)-2.
\end{equation} 
\end{thm}
This result gives us an upper bound on the rank of the Mordell-Weil lattice, hence
on the rank of the flavor group of a QFT whose universal ordinary special geometry
is given by $\mathscr{A}=W\setminus E$. If we can find an upper bound on the Picard
numbers of the $W$'s which may appear at a given rank $r$ ($\equiv\dim_\mathbb{\C(Z)}A_{\C(Z)}$)
we would have an uniform bound on the flavor symmetry valid for all SCFT of rank $r$. 

Not all fibrations $W\to Z$ whose generic fibers are dimension-$r$ Abelian varieties are allowed in special geometry.
First of all we have severe conditions on the base; when smooth $Z$ is a normal projective
compactification of $\C^k$.
In addition we have the condition that if we specialize to a generic value of the couplings $p\equiv (t_j,m_a)$
the resulting fibration $\mathscr{X}\equiv\mathscr{A}_p\to\mathscr{C}$ is Lagrangian. 
By restriction, a section of $\mathscr{A}\to\mathscr{B}$ defines a section of $\mathscr{X}\to\mathscr{C}$,
so if the couplings $p$ are sufficiently generic we have an embedding
\begin{equation}
\mathsf{MW}(\mathscr{A})\subset\mathsf{MW}(\mathscr{X}).
\end{equation}
For the purpose of getting an upper bound, we can then apply Oguiso's Shioda-Tate formula \eqref{Ofurmula} we can replace the geometry $\mathscr{A}\to\mathscr{B}$ with the
ordinary special geometry $\mathscr{X}\to\mathscr{C}$ at a \emph{generic} $p\in\mathscr{P}$. There is a pair $(M,S)$ with $M$ projective
such that $\mathscr{X}=M\setminus S$. Since $\mathscr{X}$ is symplectic, $M$
is log-symplectic, and the anti-canonical $-K_M$ is \textsf{nef.} Therefore we expect that there is an upper bound for the Picard number $\rho(M)$ of all $2r$-dimensional projective varieties which are
fibered in Abelian varieties of dimension $r$ and are log-symplectic. This upper bound minus 2 will yield an uniform upper bound for the rank of the flavor symmetry group of a rank $r$ SCFT. While finding the actual bound looks hard in general, this line of reasoning suggests that a bound on the flavor exists (as expected by physicists) and has a very deep mathematical origin. 
Without the condition of \emph{anti-affineness} of the underlying algebraic group no upper bound would arise.

\begin{exe} Consider $r=1$. In this case $M$ is an elliptic surface such that $M$ minus a fiber is
symplectic, i.e.\! such that the canonical class is minus the class of a fiber \cite{A11}.
Then $M$ is a rational elliptic surface fibered over $\mathbb{P}^1$.
The invariant entering in the Oguiso formula \eqref{Ofurmula} are
\be
\rho(W)=b_2(W)=10,\quad \rho(Z)\equiv \rho(\mathbb{P}^1)=1,\quad
\mathrm{rank}\, \mathsf{NS}(A_{\C(Z)})=1
\ee
 so that in
\begin{equation}
\mathrm{rank}\,\text{(flavor group)}\leq 10-2-\sum_i(m_i-1)\equiv 8-\sum_i(m_i-1)
\end{equation}
which is (of course) the original Shioda-Tate formula \cite{MW1}. Thus in rank $1$
we have $f\leq 8$
an inequality saturated by the Minahan-Nemeshanski $E_8$ model \cite{MN}.
\end{exe}

\subsection{Flavor symmetries. Ambiguities}

In rank 1 the theory of the flavor symmetry from the viewpoint of the Mordell-Weil
group was worked out in full detail in ref.\cite{A11}. Basically it turned out that the 
Mordell-Weil lattice $\mathsf{MWL}(A)$
 should be identified with the root lattice of the flavor group $F$ and the Ner\'on-Tate
height with the Weyl invariant pairing on the root lattice. 
However there remain some ambiguity since
these two data cannot distinguish, say, the flavor groups $SU(3)$ and $G_2$.
To eliminate all ambiguity one has to construct the actual \emph{root system}
i.e.\! describe the \emph{finitely many} elements of $\mathsf{MWL}(A)$ which are roots
of the flavor Lie algebra. This was done in \cite{A11} using the finite-subset of \emph{integral} points
in $\mathsf{MWL}(A_{\C(Z)})$
(as contrasted to the \emph{rational} ones) and some long calculations.

The root lattices and Weyl invariant pairings are expected to work in the same
way in higher dimension: they can be read directly from the Mordell-Weil lattice.
Finding the actual root system, however, goes beyond ``Diophantine geometry over
function fields''
since it requires the knowledge of  the \emph{specific} model
$\mathscr{X}$ of the Abelian variety $X_\mathbb{K}$
which describes the special geometry. In dimension 1 we had a preferred model, the N\'eron-Kodaira one
\cite{neron,MW1}, but the situation in higher dimension is not so clear.

\section{The SW differentials and the Weil correspondence}

Our proposal is that the universal special geometry is the Lagrangian fibration which describes a SCFT with all possible
couplings switched on and promoted from complex parameters
to non-propagating superfields (spurions).
From the extended special geometry one gets back the
ordinary Seiberg-Witten geometry of the theory with given masses $\underline{m}_a$ and relevant couplings
$\underline{t}_j$ by taking the symplectic quotient. In detail: one considers the level manifold
\be
i\colon M\hookrightarrow \mathscr{Y}
\ee
 where the spurions' vevs are kept fixed 
$t_j=\underline{t}_j$ and $m_a=\underline{m}_a$; $M$ is fibered over the submanifold
\be
\mathscr{C}\equiv\{(\underline{t}_1,\dots,\underline{t}_l,\underline{m}_1,\dots,\underline{m}_f,u_1,\dots,u_r)\}\subset \mathscr{B},
\ee
which we identify with the Coulomb branch $\mathscr{C}\equiv\mathsf{Spec}\,\mathbb{C}[u_1,\dots,u_r]$, 
\be
M\equiv\pi^{-1}(\mathscr{C}).
\ee
Then one takes the quotient $M/H$ where $H$ is the complex Lie group generated by the
Hamiltonian vector fields $\partial_{\alpha^j}$ and $2\pi i \,y^a\partial_{y^a}$ (cf.\! \eqref{iiiiiiqqqw}) of the Hamiltonians $t_j$ and $m_a$, respectively.
If $\varpi\colon M\to M/H$ is the natural projection, the reduced
integrable system is $(M/H, \Omega_\text{red})$ where the symplectic form  $\Omega_\text{red}$
is related to the original one $\boldsymbol{\Omega}$ in $\mathscr{Y}$ by the equation
\be\label{omred}
\varpi^*\Omega_\text{red}=i^*\boldsymbol{\Omega}.
\ee 
The fibration
$M/H\to \mathscr{C}$ is then a non-$\C^\times$-isoinvariant 
special geometry namely the integrable system of the SW geometry for the
SCFT deformed by the mass and relevant couplings $m_a=\underline{m}_a$, $t_j=\underline{t}_j$.
When these parameters are set to zero one gets back the $\C^\times$-isoinvariant geometry of the undeformed SCFT.

It remains to determine the SW differential $\lambda_\text{SW}$ for general values of the couplings.
In view of \eqref{omred} $\lambda_\text{SW}$ should be induced by the (holomorphic) universal Euler differential $\lambda_\ce$ via the symplectic
reduction on the locus $t_j=\underline{t}_j$, $m_a=\underline{m}_a$, that is, 
\be\label{omred2}
\varpi^*\lambda_\text{SW}=i^*\lambda_\ce+\text{exact}.
\ee 
Taking the exterior derivative of this equation we get back the symplectic quotient formula
\eqref{omred}. Now we compute explicitly $\lambda_\text{SW}$ case by case, and check that it has the expected properties for a physical Seiberg-Witten differential. Comparing \eqref{omred}, \eqref{omred2} we see that the differential behaves according to the Duistermaat-Heckman theorem as predicted by \cite{donagi1}.

\subsection{Torus extensions: rank-1}
We start from the very simplest case. We consider the extension of a rank-1 geometry by a $f=1$ torus.
 The mass parameter $m$ is the vev of a vector superfield which as $g_f\to0$ gets frozen to the constant value $\underline{m}$ (as determined by the the choice of boundary conditions at infinity),
and we consider the reduced geometry at a fixed value $\underline{m}$ of $m$
\begin{equation}
\mathscr{X}\equiv\mathscr{X}_{\underline{m}}\;\overset{\rm def}{=} \Big\{y\in \mathscr{Y}\colon m(y)=\underline{m}\Big\}\Big/\C^\times.
\end{equation} 
i.e.\! the ordinary rank-1 special geometry over $\mathscr{C}\simeq\C$ with mass $\underline{m}$.
We have the anti-affine extension
\begin{equation}
\xymatrix{\mathscr{Y}\ar[dr]_\Pi\ar[rr]^{\textsf{al}}&&\mathscr{A}\ar[dl]^\pi\\
&\mathscr{B}}
\end{equation}
where the generic fiber $\mathscr{A}_b$ is an elliptic curve $E_{\tau(b)}$ of period $\tau(b)$
while the generic fiber $\mathscr{Y}_b$ is an anti-affine group of the form 
$G_{w(b)}\to E_{\tau(b)}$ as in
 \textbf{Example \ref{kkkasqw12}}. Thus the extension is specified by 
two distinct sections
$s_0,s_1\colon \mathscr{B}\to \mathscr{A}$ where $s_0$ is the zero section and
$s_1$ is a generator of the rank-$1$ lattice $\Lambda$ in the Mordell-Weil group of
$\mathscr{A}$. Explicitly 
\be\label{7776xcv}
s_1\colon b\mapsto w(b)\in E_{\tau(b)},
\ee
 and $w(b)$ is not torsion for generic $b$.
We write
\begin{equation}\label{ooi90n}
\chi(z,w|\tau)\overset{\rm def}{=}\frac{\theta_3\big(\pi z-\pi w-\tfrac{1}{2}\pi -\tfrac{1}{2}\pi\tau\,\big|\, \tau\big)}{\theta_3\big(\pi z-\tfrac{1}{2}\pi -\tfrac{1}{2}\pi\tau\,\big|\, \tau\big)}
\end{equation}
and identify ($\theta\in\C$)
\begin{equation}\label{kkkas123}
G_w\leadsto \Big\{\Big(z,\chi(z,w|\tau)\,e^{i\theta}\Big)\in\C\times \C^\times\Big\}\Big/\Big(z\sim z+ m+n \tau\Big).
\end{equation}
As in $\textbf{Example \ref{kkkasqw12}}$ we write $y$ for the covering fibers' coordinate,
\begin{equation}
y\equiv \chi(z,w|\tau)\,e^{i\theta}.
\end{equation}

\smallskip

Let $U\subset \mathscr{B}$ be a sufficiently small coordinate patch with special coordinates $(m,a)$ where $m$ is a mass parameter
and $a$ is an usual local special coordinate along the Coulomb branch $\mathscr{C}$.
The restriction $\mathscr{Y}|_U$ of the extended geometry to $U$  has the form
\begin{equation}\label{hhjytjj}
\Big(m,a, z, \chi\big(z,w(m,a)|\tau(m,a)\big)\,e^{i\theta}\Big)\in U\times \C\times \C
\end{equation}
modulo the obvious identifications
\begin{gather}
z\sim z+r + s\,\tau(m,a),\quad r,s\in\Z\\
\theta\sim \theta+2\pi\, k,\quad k\in \Z.
\end{gather}
 From eq.\eqref{kkkas12zzzu} we see that the Euler differential in $\mathscr{Y}|_U$
has the local form
\begin{equation}\label{piouyt}
\lambda_\ce=m\, \frac{dy}{2\pi i\, y}+ a\,dz= m\,\frac{d\theta}{2\pi} +\frac{m}{2\pi i}\, \frac{d\chi(z,w|\tau)}{\chi(z,w|\tau)}+ a\,dz.
\end{equation}
Comparing with eq.\eqref{perper} we see that the periods of $\lambda_\ce$ are
\begin{equation}
\begin{aligned}
&\text{$A$-period:} &&a\\
&\text{$B$-period:} &&\tau a+ m w\\
&\text{$C$-period:} &&m.
\end{aligned}
\end{equation}

\subsubsection{The reduced SW differential.} 

In terms of the local coordinates in eq.\eqref{hhjytjj} the (complex)
 time evolution of the phase space $\mathscr{Y}$ generated by the Hamiltonian $m$ is
just a linear shift of the coordinate $\theta$ canonically conjugate to the momentum $m$
\begin{equation}
\theta\leadsto \theta+t,\qquad t\in \C.
\end{equation}

Hence the Legendre differential (\emph{alias} SW differential) of the reduced integrable system is obtained by simply dropping
in eq.\eqref{piouyt} the term containing the $d\theta\sim dt$. Then the SW differential
takes the local form 
\begin{equation}\label{piouyt2}
\lambda_\text{SW}=\frac{m}{2\pi i}\, \frac{d\chi(z,w|\tau)}{\chi(z,w|\tau)}+ a\,dz.
\end{equation}
The `naive' symplectic structure 
\begin{equation}\label{omla}
\Omega=d\lambda_\text{SW}=da\wedge dz\quad \text{(locally)}
\end{equation}
  is still holomorphic (as it should be in special geometry) but the SW differential
  $\lambda_\text{SW}$ is now just \emph{meromorphic}. In facts from the explicit expression \eqref{piouyt2}
  and \eqref{ooi90n} we see
that a mass-deformed \emph{ordinary} special geometry has a \emph{meromorphic} SW differential
  $\lambda_\text{SW}$ with \emph{constant} residues equal to  linear combinations of the mass parameters with integral coefficients.
 We stress that the singularities in $\lambda_\text{SW}$ are an ``artifact'' of the 
  MWM reduction procedure. The parent differential $\lambda_\ce$
  is perfectly holomorphic.

\subsubsection{The SW differential as a current}  There is a less naive point of view \cite{donagi1}.
 The proper way to understand $\lambda_\text{SW}$ is to see it as a \emph{$(1,0)$-current}
 not as a \emph{$(1,0)$-form}. The space of $(1,0)$-currents
 on a complex $n$-fold $M$ is the topological dual to the space $C^\infty_c\mspace{-1.5mu}(M)^{n-1,n}$ of compactly supported, smooth differential forms
 of type
 $(n-1,n)$  (see e.g.\! \cite{GH}).
 We write $T_\text{SW}$ for $\lambda_\text{SW}$ seen as a current i.e.\!
 \begin{equation}
 \int T_\text{SW}\, \phi=\int_M \lambda_\text{SW}\wedge \phi,\qquad \phi\in C^\infty_c\mspace{-1.5mu}(M)^{n-1,n}.
 \end{equation}
 By general theory (see page 368 of \cite{GH}) the exterior derivative of a meromorphic differential seen as a current
 differs from its `naive' differential by a `residue' term, that is, by a current with a codimension-1 support; then eq.\eqref{omla} should be replaced by
 \begin{equation}
 dT_\text{SW} =\Omega +\text{``residue''}.
 \end{equation} 
The residue is easily computed with the help of the \emph{Poincar\'e-Lelong formula.}
 
 \paragraph{The Poincar\'e-Lelong formula.} Let $D\equiv (f)\subset M$ be a smooth irreducible divisor.
 The Poincar\'e-Lelong formula is the equality of (1,1)-currents  (cf.\! page 388 in \cite{GH}) 
 \begin{equation}
 d\!\left(\frac{df}{2\pi i\,f}\right)=\delta_D
 \end{equation}
 where $\delta_D$ is the ``Dirac'' (1,1)-current defined by
 \begin{equation}
 \int \delta_D\, \phi =\int_D \phi,\qquad \phi\in C^\infty_c\mspace{-1.5mu}(M)^{n-1,n-1}.
 \end{equation}
 More generally, if $D=\sum_i a_i\, D_i$, with $D_i$ irreducible, we set
 \begin{equation}
 \delta_D=\sum_i a_i \,\delta_{D_i}.
 \end{equation}
 Clearly,  in the sense of the cohomology of currents, which is isomorphic to the usual de Rham cohomology of $M$ (see page 385 in \cite{GH}),
 we have
 \begin{equation}
 [\delta_D]\equiv [D]
 \end{equation}
 where $[D]$ is the fundamental class of $D$.
 
 \paragraph{The differential of the SW current.}
 Using the Poincar\'e-Lelong formula it is easy to see that, locally in $\mathscr{Y}|_U$, we have the equality of currents
 \begin{equation}
 dT_\text{SW}=\Omega+ m\,\delta_{(z-w(u))}-m\,\delta_{(z)}.
 \end{equation}
The corresponding global statement is
 \begin{equation}
 dT_\text{SW}= \Omega+m\Big(\delta_{s_1}-\delta_{s_0}\Big)
 \end{equation}
 where $s_0$ is the zero-section and the section $s_1$ is defined in \eqref{7776xcv}.
 Therefore, in the cohomology of currents $\equiv$ de Rham cohomology, we have 
 \begin{equation}
 [\Omega]=m \Big(\delta_{s_0}-\delta_{s_1}\Big),
 \end{equation}
as expected in \cite{donagi1}.
 
\subsection{General torus extension} 
We now extend the above result to arbitrary ranks $r$ and $f$ but still with all
relevant deformation switched off.
After the reduction to
an ordinary special geometry over the usual Coulomb branch $\mathscr{C}$, the polar locus of the 
meromorphic differential $\lambda_\text{SW}$ defines a divisor
in $\mathscr{X}$ whose
irreducible components $E_i\subset \mathscr{X}$ are complex submanifolds of dimension
$2r-1$, which are ``horizontal''  in the sense that their intersection
with each fiber, $E_i\cap \mathscr{X}_u$  defines a divisor inside the fiber $\mathscr{X}_u$. More intrinsically
they are divisors of the underlying Abelian variety $X_\mathbb{K}$ defined over the function field $\mathbb{K}$, that is,
$\mathbb{K}$-valued points of the dual Abelian variety $X_\mathbb{K}^\vee$
which is isomorphic to $X_\mathbb{K}$ (since we assume the polarization to be principal). 
The derivative
\begin{equation}
\frac{\partial}{\partial m^a}\lambda_\text{SW}
\end{equation}
is a meromorphic differential whose polar divisor $D_a$
intersects each fiber $\mathscr{X}_u$ along a 
 divisor $D_a\cap \mathscr{X}_u\subset \mathscr{X}_u$ which is required to satisfy two conditions:
 \begin{itemize}
 \item[(a)] is algebraically trivial in $\mathscr{X}_u$
 by eq.\eqref{lllk90n};
 \item[(b)] is non-torsion (in order the group to be anti-affine).
 \end{itemize}
 
 Therefore, in the general case we
 have
 \begin{gather}
 dT_\text{SW}=\Omega+\sum_a m_a\,\delta_{D_a}\\
 \Omega=-\sum_a m_a\,[D_a].\label{kasqwert33331}
 \end{gather}
 
At first sight it may seem that the situation for higher $r$
is quite different from the one we saw in rank-1, where
the polar divisor of $\lambda_\text{SW}$
is a linear combination (with integral coefficients)
of \emph{sections} of $\pi$ (submanifolds of dimension $r$)
whereas for general $r$ we have \emph{divisors} i.e.\! submanifolds of dimension
$2r-1\neq r$.   However the difference is
only apparent. Indeed, 
an algebraically trivial ``horizontal'' divisor corresponds to a $\mathbb{K}$-point in $X_\mathbb{K}^\vee$
while the principal polarization yields an isomorphism $X_\mathbb{K}^\vee\to X_\mathbb{K}$, so 
these divisors may be seen as points in $X_\mathbb{K}(\mathbb{K})$ i.e.\! sections of $\pi\colon\mathscr{X}\to\mathbb{C}$.
Let $\phi\colon X_\mathbb{K}(\mathbb{K})\to X_\mathbb{K}(\mathbb{K})^\vee$ be the isomorphism sending
 sections ($\equiv$ points in $X_\mathbb{K}(\mathbb{K})$) into ``horizontal'' divisors. One has
 \be 
 dT_\text{SW}=\Omega+\sum_a m_a\,\delta_{\phi(s_a)}
 \ee 
This formula makes clear that the mass parameters $m_a$
take value in a vector space
whose dual contains a natural 
lattice, generated by the $\delta_{\phi(s_a)}$, which carries
a natural symmetric product which is positive-definite.

\subsubsection{Relation to the Duistermaat-Heckman theorem}
This is the point emphasized by Donagi \cite{donagi1}.
The issue is to understand the deep reason
why the residues of $\lambda_\text{SW}$ are constants linear in the mass parameters $m^a$.
In view of eqn.\eqref{kasqwert33331} this amounts to asking why the
cohomology class of the symplectic form $[\Omega]$ depends linearly on the masses.

This fact follows from our construction of the mass-deformed special geometry as 
a symplectic (Hamiltonian) reduction of the extended geometry associated to the zero-coupling
gauged flavor symmetry model. The Duistermaat-Heckman theorem \cite{DH,cannas} holds for the symplectic quotient by the action of
any compact group $F$ (acting symplectically) \cite{donagi1}. For easy of exposition we take $F=U(1)$,
the generalization to arbitrary compact flavor groups being obvious (see \cite{donagi1} or
\cite{guill}).
The theorem is usually stated for real symplectic manifolds,
but everything remains valid for complex symplectic manifolds.
In this case $U(1)$ is replaced by its complexification $\C^\times\equiv GL(1,\C)$.

\medskip

We have the extended symplectic manifold $(\mathscr{Y},\boldsymbol{\Omega})$ with a Hamiltonian action of $U(1)$
generated by the Hamiltonian vector field $v$ with momentum map
\begin{equation}
\mu\colon \mathscr{Y}\to \mathfrak{gl}(1)^\vee, \quad \iota_v\boldsymbol{\Omega}=d\mu.
\end{equation}
We assume that the $U(1)$ action is free on the level set $\mathscr{Y}_{\underline{m}}\equiv\mu^{-1}(\underline{m})$. 
The reduced phase space
($\equiv$ total space of the ordinary special geometry with mass parameter $m$ set to $m_0$) is
\begin{equation}
\mathscr{X}_{\underline{m}}=\mu^{-1}(\underline{m})/GL(1,\C)
\end{equation}
so that $\mathscr{Y}_{\underline{m}}\to \mathscr{X}_{\underline{m}}$ is a principal $GL(1,\C)$ bundle. 
We fix a connection 1-form $\alpha$ on this principal bundle:
by definition one has 
\begin{equation}
\iota_v \alpha=1.
\end{equation}
By the Marsden-Weinstein-Meyer theorem,
$\mathscr{X}_{\underline{m}}$ is a symplectic manifold
with symplectic form $\omega_{\underline{m}}$.
 We consider a tubular neighborhood $T_{\underline{m}}$ of $\mathscr{Y}_{\underline{m}}$ in $\mathscr{Y}$, say
\begin{equation}
T_{\underline{m}}=\big\{y\in \mathscr{Y}\colon \underline{m}-\epsilon < \mu(y)\equiv m < \underline{m}+\epsilon\big\}\subset \mathscr{Y}.
\end{equation}
Let $\Omega_{\underline{m}}\equiv\boldsymbol{\Omega}|_{\mathscr{Y}_{\underline{m}}}$ and $\pi_{\underline{m}}\colon T_{\underline{m}}\to \mathscr{Y}_{\underline{m}}$ a projection. One checks that for small $\epsilon$ the form
\begin{equation}
\pi_{\underline{m}}^\ast\mspace{2mu} \Omega_{\underline{m}}-d((m-{\underline{m}})\,\alpha)
\end{equation}
is a symplectic form on $T_{\underline{m}}$. Hence, for $m-{\underline{m}}$ small enough,
the symplectic form on $\mathscr{X}_m\equiv \mu^{-1}(m)/GL(1,\C)$ is
\begin{equation}
\omega_m=\omega_{\underline{m}}-(m-{\underline{m}})F,
\end{equation}
where $F$ is the curvature form of the $U(1)$ connection defined by 
\begin{equation}
\pi^* F=d\alpha.
\end{equation}
This construction depends on several choices; however all choices are homotopically equivalent, 
and hence the above equation is intrinsic in cohomology  \cite{cannas}.
Therefore, passing in cohomology, we get
\begin{equation}\label{duisterrem}
[\omega_m]=[\omega_{\underline{m}}]+(m-{\underline{m}})c_1
\end{equation}
where $c_1$ is the first Chern class of the line bundle $\mathscr{Y}_{m_0}\to \mathscr{X}_{m_0}$. 
Formula \eqref{duisterrem} is the Duistermaat-Heckman theorem. We see that the class of the symplectic form
is linear in $m$, as predicted by the previous arguments.
In facts there is much more: taking the derivative of the Duistermaat-Heckman formula we get
\begin{equation}
\frac{\partial}{\partial m}[\omega_m]=c_1(\cl)
\end{equation}
where $\cl$ is the line bundle which defines the extension
of the dimension $r$ Abelian variety $A_{u,m}$ by $\C^\times$.
This is exactly the same as the derivative of eqn.\eqref{kasqwert33331}
since from our explicit construction (e.g.\! \eqref{kkkas123} for $r=1$)
we see that
\begin{equation}
\cl=\co(-D)\quad\Rightarrow\quad c_1(\cl)=-[D].
\end{equation}  

\subsection{Relevant deformations and SW differentials} The derivatives
 \begin{equation}
 \frac{\partial}{\partial t_\alpha}\lambda_\text{SW}\Big|_{X_u}
 \end{equation}
  are (generically)
meromorphic differentials of the \emph{second} kind, that is with higher order poles
and zero residues (see \cite{ati,GH}).
The relevant deformations produce higher singularities of the SW differential
than mass deformation. This agrees with the physical intuition: relevant perturbations
are \emph{less soft} than mass perturbations, so produce stronger deviations from the
superconformal situation (which corresponds to a regular $\lambda_\text{SW}$). 
This is reflected in the Weil correspondence in the fact that mass deformations lead to
``tame'' semi-stable degenerations of Abelian varieties (i.e.\! torus extensions) while
 relevant deformations lead to \emph{un}-stable degenerations.
We limit ourselves to present
a an explicit example which illustrates all the issues involved.

\begin{exe} We consider the family of extension $H_\eta$ of an elliptic curve $E$
 by the additive group $\C$
\begin{equation}
0\to \C\to H_\eta\to E \to 0
\end{equation}
which form a one-parameter family parametrized by $\eta\in\C$.
The extension group $H_\eta$ is anti-affine iff $\eta\in\C^\times$. A family of such extension groups 
describes the extended geometry of a rank-1 QFT with one relevant operator and trivial flavor group.

To pass from the multiplicative (torus extension) to the additive case (vector extension) 
we replace the automorphic \emph{factor}
we used in \textbf{Example \ref{kkkasqw12}} with the analogue automorphic \emph{summand.}
We write the elliptic curve $E$ in terms of its Weierstrass equation
\begin{equation}\label{wwwaert}
y^2=4\,x^3-g_2\,x-g_3,
\end{equation}
where $g_2$ and $g_3$ are the usual functions of the two periods $(2\omega_1,2\omega_2)$ of
the elliptic curve 
\begin{equation}
E\equiv\C/(z\sim z+2m\omega_1+2n\omega_2).
\end{equation}
The universal cover of $H_\eta$ is $\C\times \C$ whose coordinates we write in the form
(cf.\! eq.\eqref{addtive})
\begin{equation}\label{kkkkhfii9}
\big(z, x-\kappa\, \zeta(z)\big)
\end{equation}
subject to the identification
\begin{equation}
z\sim z+2m\,\omega_1+2n\,\omega_2.
\end{equation}
In eqn.\eqref{kkkkhfii9} $\zeta(z)$ is the Weierstrass $\zeta$-function which satisfies \eqref{weier}.

\medskip

Now assume that we have an extended geometry
\begin{equation}
\mathscr{Y}\to \mathscr{B}
\end{equation}
 of rank $(r,f,k)=(1,0,1)$, so that the smooth
fibers $\mathscr{Y}_b$ are groups of the form $H_{\kappa(b)}$ for some elliptic curve $E_{\tau(b)}$. 
We use standard local coordinates $(t, a)$ on $\mathscr{B}$
of dimensions $\{2-\Delta,1\}$ where $\Delta>1$ is the dimension of the \emph{global}
coordinate $u$ on the ordinary Coulomb branch $\cc$.
Then locally the canonical 
Euler differential has the form
\begin{equation}\label{jhhhasqwrt}
\lambda_\ce= a\, dz+ t\, d\big(x-\eta\, \zeta(z)\big).
\end{equation}
After the MWM reduction to the ordinary special geometry where the relevant coupling $t$ is frozen to the value $\underline{t}$, we have a SW differential defined as
\be
\lambda_\text{SW}= \lambda_\ce\bigg|_{t=\overline{t}\atop dx=0}+\text{exact}
\ee
so that
\begin{equation}
\lambda_\text{SW}= \Big(a+\overline{t}\,\eta\, \wp(z)\Big)dz.
\end{equation}
$\lambda_\text{SW}$ is now a meromorphic differential of the \emph{second kind}
\cite{ati,GH} with poles of the second order at the lattice points and zero residues in agreement with the Weil correspondence.
On the other hand the (reduced) symplectic form
\begin{equation}
\Omega\equiv d\lambda_\text{SW}= da\wedge dz,
\end{equation}
is a holomorphic $(2,0)$ form.
\end{exe}

\begin{rem} We see that when $\eta\neq0$ we can reabsorb it in the definition of
the relevant coupling $t$, so that we may set $\eta$ to a convenient constant with no loss.
\end{rem}

\subsection{Example: Argyres-Douglas QFT of type $A_2$} The above example corresponds to
an actual $\cn=2$ QFT namely the Argyres-Douglas model of type $A_2$.
It is convenient to geometrically engineer this QFT in F-theory, see e.g.\! ref.\cite{Cecotti:2010fi}.
From the F-theory construction, one gets the so-called ``SW curve'' in the form
\begin{equation}\label{wwwwertss}
y^2= 4\,x^3-t\,x-u,
\end{equation}
where $(t, u)$ are the \emph{global} coordinates in $\mathscr{B}\cong \C^2$, with $u$ the coordinate on the
ordinary Coulomb branch $\mathscr{C}$. 
The elliptic curve \eqref{wwwwertss} is the Albanese variety $A_\mathbb{L}$ of the universal special geometry of the $A_2$ SCFT, where $\mathbb{L}\equiv \C(\mathscr{B})\equiv \C(\lambda,u)$. 
The SW differential is then
\begin{equation}\label{pp12z6}
\lambda_\text{SW}\equiv y\,dx
\end{equation}
which, by definition, has dimension 1
\begin{equation}
\Delta(y\,dx)\equiv \Delta(y)+\Delta(x)=1.
\end{equation}
Comparing with the scaling of the equation \eqref{wwwwertss} we get
\begin{equation}
\Delta(y)=\frac{3}{5},\quad\Delta(x)=\frac{2}{5},\quad \Delta(t)=\frac{4}{5},\quad
\Delta(u)=\frac{6}{5}.
\end{equation}
Note that these numbers have the two properties expected from grand geometry:
$\Delta(u)$ is an allowed dimension for a rank-1 SCFT \cite{A9}
and the universal geometric relation \eqref{predphys} is satisfied:
\begin{equation}\label{hhhhhhyg}
\Delta(t)+\Delta(u)=2.
\end{equation}
The (locally defined) special coordinate $a$  is (up to a multiplicative constant)
\begin{equation}
a= u^{5/6}.
\end{equation}

From eq.\eqref{pp12z6} we read 
the SW differential on $A_\mathbb{L}$ 
\begin{equation}
\lambda_\text{SW}\equiv \sqrt{4\,x^3-t\, x-u}\,dx+\text{exact}
\end{equation}
And (neglecting the irrelevant exact term)
\begin{equation}
\begin{aligned}
\frac{\partial}{\partial a}{\lambda}_\text{SW}&=\frac{6}{5}\,u^{1/5}
\frac{\partial}{\partial u}{\lambda}_\text{SW}=
-\frac{3}{5}\,u^{1/5}\, \frac{dx}{\sqrt{4\,x^3-t\, x-u}}\\
&\equiv -\frac{3}{5}\,u^{1/5}\,\frac{dx}{y}\equiv -\frac{3}{5}\,u^{1/5}\,dz\equiv \omega+\text{exact},\\
\frac{\partial}{\partial t}{\lambda}_\text{SW}&=
\frac{1}{2}\, \frac{x\,dx}{\sqrt{4\,x^3-t\, x-u}}= \frac{1}{2}\, \wp(z)\,dz\equiv -\frac{1}{2}\, d\zeta(z),
\end{aligned}
\end{equation}
where $\omega$ is the differential of the first kind in the elliptic fiber normalized so that its periods
are $\C^\times$-invariant. 
We see that the derivative of the 
 the SW differential on the universal ordinary special geometry $A_\mathbb{L}$
 exactly matches the form \eqref{jhhhasqwrt} (with the coupling normalization 
 $\eta=1/2$) which is expected
on the grounds of the theory of anti-affine group schemes (i.e.\! from the {Weil correspondence}).
The holomorphic differential $\lambda_\ce$ on the total space of the universal geometry $
\mathscr{Y}$ is (cf.\! eq.\eqref{jhhhasqwrt})
\be
\lambda_\ce=u\, dz+t\,dy \equiv u\, dz+t\, d\!\left(x-\frac{1}{2}\zeta(z)\right).
\ee
While the ordinary special geometry is not $\C^\times$-isoinvariant when $\underline{t}\neq0$,
the universal one is still $\C^\times$-isoinvariant. This amounts to saying that the period $\tau(t,u)$ 
of the Albanese variety of the
fiber $\mathscr{Y}_{(t,u)}$ over $(t,u)\in\mathscr{B}$ depends only on the scale invariant combination
\begin{equation}
v= u/t^{3/2},\qquad \mathscr{L}_\ce\,v=0\quad\Rightarrow\mathscr{L}_\ce\, \tau=0.
\end{equation}
This is a well-known property of the elliptic curve $A_\mathbb{L}$ in eq.\eqref{wwwwertss}.


\begin{thebibliography}{183}
  \begin{small}
  

\bibitem{SW1}
  N. Seiberg and E. Witten, Electric-magnetic duality, monopole condensation, and confinement in N=2 supersymmetric Yang-Mills theory, Nucl. Phys. B 426, 19 (1994) Erratum: Nucl. Phys. B 430, 485 (1994), arXiv:hep-th/9407087.

 

  \bibitem{SW2}
N. Seiberg and E. Witten, Monopoles, duality and chiral symmetry breaking in N=2 super- symmetric QCD, Nucl. Phys. B 431, 484 (1994), arXiv:hep-th/9408099.

 
 \bibitem{donagi0}
R. Donagi and E. Witten, Supersymmetric Yang-Mills theory and integrable systems,
Nucl. Phys. B 460, 299 (1996), arXiv:hep-th/9510101.

 
 \bibitem{donagi1}
R. Donagi, Seiberg-Witten integrable systems, in \textit{Surveys in Differential Geometry}, Vol. IV (1998) pp. 83-129.




\bibitem{A1}
P. C. Argyres, M. Crescimanno, A. D. Shapere and J. R. Wittig, Classification of N=2 superconformal field theories with two-dimensional Coulomb branches, hep-th/0504070.

\bibitem{A2} P. C. Argyres and J. R. Wittig, Classification of N=2 superconformal field theories with two-dimensional Coulomb branches. II, hep-th/0510226.

\bibitem{A3}
P. Argyres, M. Lotito, Y. L\"u and M. Martone, Geometric constraints on the space of N=2 SCFTs I: physical constraints on relevant deformations, arXiv:1505.04814 [hep-th].

\bibitem{A4}
P. C. Argyres, M. Lotito, Y. L\"u and M. Martone, Geometric constraints on the space of N=2 SCFTs II: Construction of special Ka\"hler geometries and RG flows, JHEP 1802, 002 (2018), arXiv:1601.00011 [hep-th].

\bibitem{A5}
 P. C. Argyres, M. Lotito, Y. L\"u and M. Martone, Expanding the landscape of N = 2 rank 1 SCFTs, JHEP 1605, 088 (2016) [arXiv:1602.02764 [hep-th]].
 
 
\bibitem{A6}
 P. Argyres, M. Lotito, Y. L\"u and M. Martone, Geometric constraints on the space of N=2 SCFTs III: enhanced Coulomb branches and central charges, arXiv:1609.04404 [hep-th].
 
 \bibitem{A7}
 P. C. Argyres and M. Martone, 4d N =2 theories with disconnected gauge groups, JHEP
1703, 145 (2017)arXiv:1611.08602.
 
 \bibitem{A8}
P. C. Argyres, Y. L\"u and M. Martone, Seiberg-Witten geometries for Coulomb branch chiral rings which are not freely generated, JHEP 1706, 144 (2017) [arXiv:1704.05110 [hep-th]].

 \bibitem{A9}
M.~Caorsi and S.~Cecotti,
Geometric classification of 4d $\mathcal{N}=2$ SCFTs,
JHEP \textbf{07}, 138 (2018),
arXiv:1801.04542.

\bibitem{A10}
P. C. Argyres, C. Long and M. Martone, The singularity structure of scale-invariant
rank-2 Coulomb branches, arXiv:1801.01122 [hep-th].


\bibitem{A11}
M.~Caorsi and S.~Cecotti,
Special Arithmetic of Flavor,
JHEP \textbf{08}, 057 (2018)
[arXiv:1803.00531 [hep-th]].


\bibitem{A12}
P.~C.~Argyres and M.~Martone,
Coulomb branches with complex singularities,
JHEP \textbf{06}, 045 (2018),
arXiv:1804.03152.


\bibitem{A13}
M.~Caorsi and S.~Cecotti,
Homological classification of 4d $\mathcal{N}$ = 2 QFT. Rank-1 revisited,''
JHEP \textbf{10} (2019), 013,
arXiv:1906.03912.


\bibitem{A14}
P.~Argyres and M.~Martone,
Construction and classification of Coulomb branch geometries,
arXiv:2003.04954.

\bibitem{A15}
M.~Martone,
Towards the classification of rank-r $\mathcal{N} $ = 2 SCFTs. Part I. Twisted partition function and central charge formulae,
JHEP \textbf{12}, 021 (2020),
arXiv:2006.16255.



\bibitem{A16}
P.C. Argyres and M. Martone,
Towards a classification of rank $r$ $N=2$ SCFTs Part II: special Kahler stratification of the Coulomb branch,
arXiv:2007.00012.

\bibitem{A17}
P.~C.~Argyres and M.~Martone,
The rank 2 classification problem I: scale invariant geometries,
arXiv:2209.09248.

\bibitem{A18}
P.~C.~Argyres and M.~Martone,
The rank 2 classification problem II: mapping scale-invariant solutions to SCFTs,
arXiv:2209.09911.

\bibitem{A19}
P.~C.~Argyres and M.~Martone,
The rank-2 classification problem III: curves with additional automorphisms,
arXiv:2209.10555.


\bibitem{A20}
M.~Martone,
The constraining power of Coulomb Branch Geometry: lectures on Seiberg-Witten theory,
arXiv:2006.14038.

\bibitem{A21}
S.~Cecotti, M.~Del Zotto, M.~Martone and R.~Moscrop,
The characteristic dimension of four-dimensional ${\mathcal {N}}$~=~2 SCFTs,''
Commun. Math. Phys. \textbf{400}, no.1, 519-540 (2023),
arXiv:2108.10884 [hep-th].

\bibitem{A22}
S.~Cecotti,
Hwang-Oguiso invariants and frozen singularities in special geometry,
arXiv:2304.04481.

\bibitem{A23}
S.~Cecotti,
Direct and Inverse Problems in Special Geometry,
[arXiv:2312.02536 [hep-th]].

\bibitem{DH}
J.J. Duistermaat and G. Heckman, On the variation in the cohomology of the symplectic
form on the reduced phase space, Invent. Math. \textbf{69} (1982) 259-268; Addendum:
Invent. Math. \textbf{72} (1983) 153-158. 

\bibitem{cannas}
A. Cannas da Silva, \textit{Lectures on symplectic geometry,}
Lecture Notes in Mathematics \textbf{1764} Springer 2008.

\bibitem{milne}
J.S. Milne, \textit{Algebraic Groups. The theory if group schemes of finite type over a field,}
Cambridge studies in advanced mathematics \textbf{170}, CUP 2017.

\bibitem{MM}
B. Mazur and W. Messing, \textit{Universal Extensions and One Dimensional Crystalline
Cohomology,} Lecture Notes in Mathematics 370, Springer 1974.

\bibitem{Weil}
A. Weil, \textit{Vari\'eti\'es ab\'eliennes,} Colloque d'alg\`ebre er theories des nombres,
Paris, 1949.

\bibitem{barsotti}
I. Barsotti, Factor sets and differentials on Abelian varieties, Trans. AMS \textbf{84} (1957)
85-108.

\bibitem{serre1}
J.-P. Serre, Expos\'e XI in S\'eminaire Chevalley 1958-59,
Varieties de Picard. Secretariat Mathematique, Paris (1960).

\bibitem{serre2}
J.-P. Serre, \textit{Algebraic Groups and Class Fields,}
 Graduate Texts in Mathematics 117, Springer (1988).

\bibitem{ttstar}
S.~Cecotti and C.~Vafa,
Topological antitopological fusion,
Nucl. Phys. B \textbf{367} (1991), 359-461

\bibitem{ttf1}
E. Witten, Topological quantum field theory, Comm. Math. Phys. \textbf{117} (1988) 353-386.



\bibitem{witten}
E. Witten, Quantum Field Theory and the Jones Polynomial, Commun. Math. Phys. \textbf{121} 351 (1989).

\bibitem{marino}
J. Labastida and M. Mari\~no, \textit{Topological Quantum Field Theory and Four Manifolds,}
Springer, 2005.

\bibitem{wittenM}
E. Witten, Monopoles and four manifolds, Math. Res. Lett. \textbf{1} (1994) 769 [hep-
th/9411102].

\bibitem{gaiotto2}
D. Gaiotto, G. W. Moore, and A. Neitzke, Four-dimensional wall-crossing via three-dimensional field theory, arXiv:0807.4723.

  
\bibitem{trace1}
W. L. Chow, Abelian varieties over function fields, Trans. AMS \textbf{78} (1955)  253-275.

\bibitem{trace2}
W. L. Chow, On abelian varieties over function fields, Proc. Natl. Acad. Sci. USA  \textbf{41} (1955) 582-586.


\bibitem{langII}
S. Lang, \textit{Fundamentals of Diophantine Geometry,} 
Springer 1983.
 

\bibitem{langIII}
S. Lang, \textit{Number Theory III,} Encyclopaedia of Mathematical Sciences, vol. 60
Springer 1991.
  

\bibitem{langI}
S. Lang, \textit{Abelian Varieties,}
Springer 1983.

\bibitem{milne2}
J.S. Milne, \textit{Abelian varieties,}
notes available at\\ \texttt{https://www.jmilne.org/math}. 

  


  
  
  
\bibitem{brion}
M. Brion, Anti-affine algebraic groups, {\tt arXiv:0710.5211}.

\bibitem{sancho1}
C. Sancho de Salas, \textit{Grupos algebraicos y teor\'ia de invariantes,}
Aportaciones Matem\'aticas: Textos \textbf{16} Sociedad Matem\'atica Mexicana 2001
(in spanish).

\bibitem{sancho2}
C. Sancho de Salas and F. Sancho de Salas, Principal bundles, quasi-abelian varieties and structure of algebraic groups
{\tt arXiv:0806.3712}.

 \bibitem{dlmf}
 \textit{NIST Digital Library of Mathematical Functions,}
 available on-line at {\tt https://dlmf.nist.gov}.
 
 \bibitem{ros58}
M. Rosenlicht, Extensions of vector groups by abelian varieties,
Amer. J. Math. \textbf{80} (1958) 685-714.

\bibitem{MW1}
M. Sch\"utt and T. Shioda, \textit{Mordell–Weil Lattices,}
A Series of Modern Surveys in Mathematics 70, Springer (2019)






\bibitem{NL}
S. Lang and A. N\'eron,
Rational points of abelian varieties over function fields, Amer. J. Math. \textbf{81} (1959) 95-118.

\bibitem{Oguiso}
K. Oguiso, Shioda-Tate formula for an Abelian fibered variety and applications,
{\tt arXiv:math/0703245.}

\bibitem{ST2}
B. Kahn, D\'emonstration g\'eom\'etrique du Th\'eor\'eme de Lang-N\'eron, {\tt arXiv:math.AG/0703063.}

\bibitem{ST3}
M. Hindry, A. Pacheco, and R. Wazir, Fibrations et conjecture de Tate, J. Number Theory \textbf{112} (2005) 345--368.

\bibitem{MN}
J. A. Minahan and D. Nemeschansky, Superconformal fixed points with E(n) global symmetry, Nucl. Phys. \textbf{B 489} (1997) 24  [hep-th/9610076].

\bibitem{neron}
S. Bosch, W. L\"utkebohmert, M. Raynaud, \textit{N\'eron Models,} Ergebnisse der Mathematik und ihrer Grenzgebiete, vol. 21 (Springer, Berlin, 1990)


\bibitem{GH}
P. Griffiths and J. Harris,
\textit{Principles of Algebraic Geometry}, Wiley (1978).


\bibitem{guill}
V. Guillemin, \textit{Moment maps and combinatorial invariants of Hamiltonian $T^n$-spaces,}
Progress in Mathematics \textbf{122} Birkh\"auser 1994. 


\bibitem{ati}
M.F. Atiyah and W.V.D. Hodge,
Integrals of the second kind on an algebraic variety, Ann. of Math. \textbf{62} (1955) 56-91.

  \bibitem{Cecotti:2010fi}
S.~Cecotti, A.~Neitzke and C.~Vafa,
R-Twisting and 4d/2d Correspondences,
{\tt [arXiv:1006.3435 [hep-th]].}

  \end{small}
  \end{thebibliography}
\end{document}